\documentclass[a4paper,11pt,preprintnumbers]{article}
\pdfoutput=1 

\usepackage{jheppub} 
\usepackage[T1]{fontenc} 
\usepackage{lmodern,bbm}

\usepackage{graphicx}
\usepackage{xspace}
\usepackage[small]{subfigure}
\usepackage{color}
\usepackage{multirow}
\usepackage{marginnote}
\usepackage{slashed}

\newcommand{\nn}{\nonumber} 
\newcommand{\bn}{{\bar n}}

\newcommand{\mcdot}{\!\cdot\!}

\newcommand{\be}{\begin{equation}}
\newcommand{\ee}{\end{equation}}
\newcommand{\bes}{\begin{subequations} \begin{align} }
\newcommand{\ees}{\end{subequations}\end{align} }
\newcommand{\bea}{\begin{eqnarray}}
\newcommand{\eea}{\end{eqnarray}}

\newcommand{\SCETa}{\mbox{${\rm SCET}_{\rm I}$ }}
\newcommand{\SCETb}{\mbox{${\rm SCET}_{\rm II}$ }}

\newcommand{\vect}[1]{\mathbf{#1}}

\newcommand{\bra}[1]{\left\langle #1\right\rvert}
\newcommand{\ket}[1]{\left\lvert #1\right\rangle}

\newcommand{\Lqcd}{\Lambda_{\text{QCD}}}

\newcommand{\GeV}{\text{ GeV}}

\newcommand{\as}{\alpha_s}

\newcommand{\cusp}{\mathrm{cusp}}

\newcommand{\wt}[1]{\widetilde{#1}}
\newcommand{\ta}{\tau_a}

\newcommand{\tcI}{{\widetilde{\mathcal{I}}}}

\newcommand{\cJ}{\mathcal{J}}
\newcommand{\cO}{\mathcal{O}}

\newcommand{\cI}{\mathcal{I}}

\newcommand{\cH}{\mathcal{H}}

\newcommand{\cL}{\mathcal{L}}

\newcommand{\cB}{\mathcal{B}}
\newcommand{\cP}{\mathcal{P}}

\newcommand{\e}{\epsilon}

\newcommand{\tS}{{\widetilde S}}
\newcommand{\tJ}{{\widetilde J}}

\newcommand{\tcB}{{\widetilde {\mathcal{B}}}}

\newcommand{\eq}[1]{Eq.~\eqref{eq:#1}}
\newcommand{\eqs}[2]{Eqs.~\eqref{eq:#1} and \eqref{eq:#2}}
\newcommand{\eqss}[3]{Eqs.~\eqref{eq:#1}, \eqref{eq:#2}, and \eqref{eq:#3}}
\renewcommand{\sec}[1]{Sec.~\ref{sec:#1}}

\newcommand{\appx}[1]{App.~\ref{app:#1}}
\newcommand{\fig}[1]{Fig.~\ref{fig:#1}}

\newcommand{\tab}[1]{Table~\ref{tab:#1}}

\newcommand{\qB}{q_B}
\newcommand{\qJ}{q_J}

\newcommand{\nB}{n_B}
\newcommand{\nJ}{n_J}

\newcommand{\bnB}{{\bar n}_B}
\newcommand{\bnJ}{{\bar n}_J}

\newcommand{\comment}[1]{}

\newcommand{\fo}{\text{fixed}} 
\newcommand{\fraca}{\frac{2(1-a)}{2-a} }
\newcommand{\wtc}{\widetilde{c}}

\allowdisplaybreaks[1]

\begin{document}

\today
\title{Angularity in DIS at next-to-next-to-leading log accuracy}

\author[a,*]{Jiawei Zhu,
\note[*]{These authors contributed equally to this work.}}
\author[a,*]{Daekyoung Kang,}
\author[a]{Tanmay Maji}

\affiliation[a]{Key Laboratory of Nuclear Physics and Ion-beam Application (MOE) and Institute of Modern Physics,
Fudan University, Shanghai, China 200433}

\emailAdd{dkang@fudan.edu.cn}
\emailAdd{tanmay@fudan.edu.cn}
\emailAdd{zhujw19@fudan.edu.cn}

\abstract{
Angularity is a class of event-shape observables that can be measured in deep-inelastic scattering. With its continuous parameter $a$ one can interpolate angularity between thrust and broadening and further access beyond the region.
Providing such systematic way to access various observables makes angularity attractive in analysis with event shapes. We give the definition of angularity for DIS and factorize the cross section by using soft-collinear effective theory. The factorization is valid in a wide range of $a$ below and above thrust region but invalid in broadening limit. It contains an angularity beam function, which is new result and we give the expression at $\mathcal{O}(\as)$. We also perform large log resummation of angularity and make predictions at various values of $a$ at next-to-next-to-leading log accuracy.}

\maketitle

\clearpage

\section{Introduction}\label{sec:intro}
Deep inelastic scattering (DIS) \cite{Bjorken:1968dy,Taylor:1991ew,Kendall:1991np,Friedman:1991nq,DIS} is an important tool to probe internal structure of nucleus and to test our understanding of Quantum Chromodynamics (QCD) \cite{Gross:1973id,Politzer:1973fx}. Newly proposed future electron-ion collider (EIC) \cite{AbdulKhalek:2021gbh,Accardi:2012qut} and  electron-ion collider in China (EicC) \cite{Anderle:2021wcy} would provide another opportunities to test and to improve our understanding of the structure of nuclei as well as dynamics of QCD in collisions.

One of classic ways of studying QCD events is to measure event shape variables that are observables designed to characterize the geometric shape of hadron distribution in the event. Their values can tell us if the distribution is pencil-like, planar, and spherical. Thrust \cite{Farhi:1977sg} characterising dijet events is one of classic event shapes predicted to very high accuracy 
\cite{GehrmannDeRidder:2007bj,GehrmannDeRidder:2007hr,Weinzierl:2008iv,Weinzierl:2009ms,Becher:2008cf,Abbate:2010xh,Mateu:2012nk} in $e^+e^-$ collisions and analysis with thrust determined the strong coupling constant $\as$ at 1\% level precision \cite{Becher:2008cf,Abbate:2010xh}.
The shape variables are also studied at hadron collisions to probe jet substructures \cite{Salam:2009jx,Ellis:2010rwa,Altheimer:2012mn,Larkoski:2017jix,Procura:2018zpn,Kang:2018qra,Caletti:2021oor,ATLAS:2019kwg,ALICE:2021njq}. They were also extensively studied in DIS \cite{Dasgupta:2003iq}. 
Improved theoretical predictions \cite{Kang:2013nha,Kang:2013lga} and new predictions \cite{Aschenauer:2019uex,Li:2020bub,Kang:2020fka,Li:2021txc} for event shapes that were not measured before should be valuable for future experiments such as EIC and EicC. 
Angularity \cite{Berger:2003iw} compared to other event shapes was defined relatively later and was not measured in DIS process. One of its features is that the angularity is rather a class of observables defined by a continuous parameter $a$, which puts weights on rapidity factor and changes relative contributions of particles with large rapidity compared to ones with smaller rapidity. The definition of angularity for a hadronic group $X$ can be written as
\bea \label{eq:taX}
\ta(X) = \frac{1}{Q_N} \sum_{i\in X} |p^i_\perp| e^{-\eta_i(1-a)} \,,
\eea
where the sum is over all hadrons $i$ in the group $X$,  $p_\perp^i$ and $\eta_i$ are transverse momentum and rapidity defined with respect to a given axis.  $Q_N$ is usually chosen to be a hard momentum scale of the system.
The angularity parameter $a$ can be any real number in region $-\infty<a<2$.  
It is known to be infrared-unsafe for $a\ge 2$ since particles are weighted rapidity too strongly and the angularity becomes sensitive to collinear splitting. \cite{Hornig:2009vb}.
Under change of $a$, the value of $\ta$ is sensitive for the collinear with large rapidity  while it is less sensitive for the soft with relatively smaller rapidity. So, one can control relative contribution between the collinear and soft with the value of $a$.
The angularity reduces to well-known event shapes: thrust when $a=0$ and broadening when $a=1$. 
Factorization in soft-collinear effective theory (SCET) \cite{Bauer:2000ew,Bauer:2000yr,Bauer:2001ct,Bauer:2001yt,Bauer:2002nz} near and below thrust region $a<1$ was known a while ago \cite{Bauer:2008dt,Hornig:2009vb}
while broadening region is understood \cite{Budhraja:2019mcz} more recently. 

Angularity axis was originally chosen to be the thrust axis \cite{Berger:2003iw} in $e^+e^-$ annihilation and the axis and its back-to-back make hadrons grouped into two-hemisphere regions, left  $\cH_L$ and right $\cH_R$, which sum up to give the total angularity $\ta^{ee}=\ta(\cH_L)+\ta(\cH_R)$.\footnote{Note that unlike \cite{Berger:2003iw,Hornig:2009vb} we do not take the absolute value $|\eta_i|$ in \eq{taX} because it is assumed that particles are already grouped into $X$ by some algorithm and their rapidity in the region is positive.} In jet substructure studies, jet axis and its constituents are usually defined by jet algorithms. Variants of angularity with alternative axis choices \cite{Larkoski:2014uqa} can be made out of different purposes. 

In this paper, we study DIS angularity defined by two axes, beam ($B$) axis along proton-beam direction and jet ($J$) axis associated with a leading jet in the final state as in \cite{Kang:2013nha}. For the definition of jet axis, we consider 1-jettiness axis that minimizes the DIS 1-jettiness or, jet axis defined by typical jet algorithms such as cone, C/A, $k_t$ and anti-$k_t$. Both axes are equivalent at leading power in small $\ta$ expansion and the axis difference is suppressed by powers of $\ta$.
Of course, one may consider alternative axes such as $z$-axis in the Breit or, CM frame and broadening axis \cite{Larkoski:2014uqa}, even multiple axes associated with multi-jet angularity. For a simplicity of discussion we focus on our research on two axes with proposed axis. We measure  angularity of a global event obtains contributions from two regions \footnote{Here, the definition of DIS angularity includes both hemispheres of the event hence, is directly sensitive to initial state radiations in beam region as well as final-state radiations in jet region. Our observable and its factorization are distinguished from jet angularity \cite{Aschenauer:2019uex}, which is defined from jet constituents being sensitive to final state radiations while insensitive to initial state radiations upon neglecting corrections of jet radius and effect of non-global logarithms.}  
\bea \label{eq:taDIS}
\ta^\text{DIS} =  \ta(\cH_B)+\ta(\cH_J) \,,
\eea
where $\cH_B$ and $\cH_J$ are beam and jet hemispheres each of which has respective jet and beam axes. The events are populated in small angularity region associated with soft and collinear radiations. We study this region with leading-power factorization by using SCET in $a<1$ region and make the prediction for  $\ta^\text{DIS}$ at next-to-next-leading log (NNLL) accuracy.

The angularity was measured in $e^+e^-$ annihilation by LEP \cite{Achard:2011zz}. It was not measured in DIS while its simplified versions thrust and broadening were measured and analyzed at HERA by the ZEUS and H1 collaborations \cite{Adloff:1997gq,Adloff:1999gn,Aktas:2005tz,Breitweg:1997ug,Chekanov:2002xk,Chekanov:2006hv}. Like other event shapes one can use the measurement to determine the strong coupling constant. Having multiple observables at different values of $a$ would be beneficial in studying correlations between the value of coupling and non-perturbative effect systematically as discussed in \cite{clee:SCET2021}. For the given tension in the values of the strong coupling constant determined from the thrust \cite{Abbate:2010xh} and $C$-parameter \cite{Hoang:2014wka} deviated several standard deviations away from the world average significantly weighted by Lattice determinations (FLAG2019) \cite{Zyla:2020zbs,Luisoni:2020efy}. New determination with the angularity from independent process such as DIS provides independent path to address this issue.

Our paper is organized as follows.
We first give our definition of axes and express the angularity in terms of Lorentz invariant four-vector products in \sec{ang}, show the factorization of angularity cross section in terms of hard, beam, jet, and soft functions derived by using SCET in \sec{factorization}, then resum logarithms of $\ta$ in Laplace space in \sec{resum}. Our  angularity beam function at $\mathcal{O}(\as)$ is given in \sec{beamfunc} and numerical results for resummed cross section at NNLL accuracy is given in  \sec{numerical}. Finally, we conclude in \sec{con}.

\section{Angularity in DIS} \label{sec:ang}
In this section, we first review kinematic variables of DIS, then give the definition of DIS angularity with the beam and jet axes, finally express the angularity in term of four-vector dot products so that it remains unchanged in any frame under arbitrary boost.

In DIS, an electron with momentum $k$ scatters off a proton of momentum $P$ by exchanging a virtual photon with a large momentum transfer $ q $. The space-like photon momentum can be written by a positive definite quantity $Q^2=-q^2$, where $ Q$ sets the momentum scale of the scattering $Q \gg \Lambda_{QCD}$.   Bj\"{o}rken scaling variable $x=Q^2/(2 P\mcdot q)$ ranges between $0 \leq x \leq 1$ and inelasticity  $y=Q^2/(x s)$ ranges between $0 \leq y \leq 1$ where $s=2k\mcdot P$ is the virtuality of electron and proton system.  

Jet productions are dominated by tree level process, where incoming quark of momentum  $xP$ from the proton is struck by the photon and propagates into the final quark with momentum $xP+q$.  The jet momentum $\vect{P}_\text{jet}$ should be close to the quark momentum but not the same due to soft radiations induces recoil and change transverse momentum of the jet.\footnote{The Breit frame makes the picture clear since photon and proton are aligned along z-axis: $P=\tfrac{Q}{x}\tfrac{\bar{n}_z}{2}$ and $q=Q (\tfrac{n_z}{2}-\tfrac{\bar{n}_z}{2})$  where $n_z=(1,0,0,1)$ and $\bar{n}_z=(1,0,0,-1)$. Then, $q_B$ becomes $Q\tfrac{\bar{n}_z}{2}$ and $q_J$ is close to $xP+q = Q\tfrac{n_z}{2}$ and they are close to back-to-back axis.}
\be
q_B=x P \,,\quad\quad 
q_J=(|\vect{P}_\text{jet}|\,,\vect{P}_\text{jet})
\,,\label{eq:qBqJ}\ee
where the axis momenta are massless $q_B^2=q_J^2=0$.  A particle momentum of $p$ can be expressed in terms of these axes and transverse momentum orthogonal to these axes\footnote{For back-to-back unit vector $n=(1,\hat n)$ and $\bar{n}=(1,-\hat n)$ with $n\mcdot \bar n=2$, 
$p=n\mcdot p \frac{\bar n}{2}+\bar n \mcdot p \frac{n}{2}+p_\perp$.}
\bea 
p=q_J\mcdot p \frac{q_B}{q_B\mcdot q_J} + q_B\mcdot p \frac{q_J}{q_B\mcdot q_J} +p_\perp
\,,
\label{eq:p}\eea
where $p_\perp\mcdot q_J=p_\perp\mcdot q_B=0$. The magnitude of transverse momentum is
\bea
|p_\perp|^2=\frac{2\, q_J\mcdot p\, q_B\mcdot p}{q_B\mcdot q_J} 
\,.
\label{eq:pperp2}\eea
We define a generalized rapidity by using $q_B$ and $q_J$, which is not back-to-back
\bea
\eta_{BJ}=-\frac12 \ln{\frac{q_B\mcdot p}{q_J\mcdot p}}
\,,
 \label{eq:hBJ}\eea
and its conjugate rapidity $\eta_{JB}=-\eta_{BJ}$ has an opposite sign. If the $p$ is closer to beam axis, the value of $\eta_{BJ}$ is positive and if it is closer to jet axis, it is negative. This defines beam and jet hemispheres: 
\bea
p\in \cH_B\quad \text{for}\quad \eta_{BJ}>0
\qquad \text{and} \qquad 
p\in\cH_J\quad \text{for}\quad \eta_{BJ}<0
\,.
\label{eq:HBJ}\eea
One can define beam angularity $\ta^B(p)$ by replacing $\eta$ with $\eta_{BJ}$ in \eq{taX} and jet angularity $\ta^J(p)$ with $\eta_{JB}$. We take $\ta^B$ when $p\in \cH_B$ and $\ta^J$ when $p\in \cH_J$: $\ta^B(p)\theta(\eta_{BJ})+\ta^J(p)\theta(\eta_{JB})$. This is equivalent to the angularity with absolute rapidity $|\eta_{BJ}|$ for all particles in entire regions as in $e^+e^-$ angularity. We can also use min/max operator and we take smaller one like $\min\{\ta^B, \ta^J \}$ for $a<1$
but larger one like $\max\{\ta^B, \ta^J \}$ for $a>1$. 

Inserting the transverse momentum in \eq{pperp2} and rapidity in \eq{hBJ} into \eq{taX} gives the angularity in term of Lorentz scalars. 
We choose $Q_N=Q^2/\sqrt{2 q_B\mcdot q_J}$ in \eq{taX} that reduces to the definition of 1-jettiness \cite{Kang:2013nha} in $a\to 0$ limit. Applying this for global event, DIS angularity is expressed as
\bea  
\ta = \frac{2}{Q^2} \sum_{i \epsilon \mathcal{H}_B} (q_B\mcdot p_i)\bigg(\frac{q_B.p_i}{q_J.p_i} \bigg)^{-a/2} + \frac{2}{Q^2} \sum_{i \epsilon \mathcal{H}_J} (q_J.p_i)\bigg(\frac{q_J.p_i}{q_B.p_i} \bigg)^{-a/2} 
\,.
\label{eq:tau_df}\eea

\section{Factorized cross section} \label{sec:factorization}
The full QCD cross section for inclusive DIS is conventionally organized into leptonic and hadronic tensors. Here, the hadronic tensor is defined with an additional measurement of angularity and the angularity cross-section for inclusive DIS is written as
\bea
\frac{d \sigma}{dx d Q^2 d \tau_a} = L_{\mu \nu}(x,Q^2)~ W^{\mu \nu}(x,Q^2,\tau_a)
\label{eq:sigdef}\,,
\eea
where the leptonic tensor is given by 
\bea 
L_{\mu\nu}(x,Q^2)=\frac{\alpha_\text{em}^2 Q_f^2}{2x^2 s^2}
\left(-g_{\mu\nu}+2\frac{k_\mu k'_\nu+k'_\mu k_\nu}{Q^2} \right)
\,,\label{eq:lepdef}\eea
where $\alpha_\text{em}$ is the fine-structure constant of QED and $Q_f$ is the electric charge of a quark with flavor $f$.
The hadronic tensor defined by QCD current $J^\mu(x)=\bar{\psi} \gamma^\mu \psi(x)$ with a delta function that measures $\ta$ as 
\bea
W^{\mu \nu}(x,Q^2,\tau_a) &=& \sum_X \langle P| J^{\mu \dagger} |X \rangle \langle X|J^\nu | P \rangle (2\pi)^{(4)} \delta^4(P+q-p_X) \delta(\tau_a-\tau_a(X))
\nn\\
 &=& \int d^4x \, e^{iq\cdot x} \langle P| J^{\mu \dagger}(x) \delta(\tau_a-\hat{\tau}_a) J^\nu (0) | P \rangle \,. 
\label{eq:Wadef}\eea
In second line we removed the sum over $X$ and replace the measure $\ta(X)$ by an operator $\hat{\tau}_a$ which acts on state $X$ as
\be 
\hat{\tau}_a |X\rangle = \tau_a(X)|X\rangle.
\label{eq:hatta}\ee
The operator $\hat{\tau}_a$ can be written in terms of momentum flow operators as discussed in \cite{Bauer:2008dt} and we bravely omit the discussion by referring to \cite{Bauer:2008dt}.

Now we discuss power counting with a small parameter $\lambda$ and factorization of angularity in the framework of SCET \cite{Bauer:2000ew,Bauer:2000yr,Bauer:2001ct,Bauer:2001yt,Bauer:2002nz}. The parameter $\lambda$ characterizes the scale of collinear and soft momenta constrained by an observable here $\tau_a$ or, some cutoffs. To describe momentum scale of soft and collinear modes in light-cone coordinates along beam and jet axes, we define light-like unit vectors $n_{B,J}$
\bea 
q_B^\mu= \omega_B \frac{n_B^\mu}{2}
\qquad
\text{and}
\qquad
q_J^\mu= \omega_J \frac{n_J^\mu}{2}
\,,
\label{eq:nBJ}\eea
where $n_{i}=(1,\hat{n}_{i})$ with $i=\{B,J\}$ and $\hat{n}_{i}$ are unit vectors along beam/jet directions defined in \eq{qBqJ}, respectively. Conjugate vectors $\bar{n}_i$ with normalization $n_i\cdot \bar{n}_i=2$ can be defined as 
\be
\bnB^\mu= \frac{2}{n_B\cdot n_J}n_J^\mu
\quad \text{and} \quad
\bnJ^\mu= \frac{2}{n_B\cdot n_J}n_B^\mu
\,.
\label{eq:bnBJ}\ee
Then, we obtain $\omega_B=\bnB\mcdot q_B$ and $\omega_J=\bnJ\mcdot q_J$.

In the beam region a particle of momentum $p$ can be expressed in terms of $n_B$ and $\bar{n}_B$: $p=p^+ \tfrac{\bar{n}_B}{2}+p^-\tfrac{n_B}{2}+p_\perp$ or,
$p=(p^+,p^-,p_\perp)=(n_B\mcdot p,\bar{n}_B\mcdot p,p_\perp)$ and in the same way for the jet region. Collinear and soft modes $p_{c,s}$ and angularity $\ta^B(p_{c,s})$ of each mode in the beam region can be expressed as
\bea 
p_c &\sim& Q(\lambda_c^2, 1, \lambda_c)\,, \,\qquad 
	\tau_a^B (p_c)\sim \lambda_c^{2-a}
\nn\\
p_s &\sim& Q(\lambda_s, \lambda_s, \lambda_s)\,, \qquad 
	\tau_a^B (p_s)\sim \lambda_s
\label{eq:pcps}\eea
This implies relevant soft mode contributing to $\tau_a^B$ has a scale of 
\be
\lambda_s \sim \lambda_c ^{2-a}
\,.
\label{eq:lamcs}\ee
We take $\lambda_c$ as a main power counting parameter and simply denote $\lambda$ by dropping the subscript $c$  from now on.
Power counting of soft mode changes with $a$.
In the limit $a\to 1$, virtualities of soft and collinear modes are of the same order $p_c^2\sim ~p_s^2$ and so is their  transverse momenta which makes recoil effect of soft radiation important in $\SCETb$. For $a$ below and away from the limit ($a<1$), virtuality and transverse momentum of soft mode are parametrically suppressed and this is so called ultra-soft mode, which makes $\SCETa$ factorization insensitive to the recoil effect. Here, we pay our attention to $\SCETa$ region.
With this observation one can express the collinear momentum as $p_c=\tilde{p}+k$, where the label momentum $\tilde{p}=\bar{n}_B\cdot \tilde{p} \, \tfrac{n_B}{2}+\tilde{p}_\perp$ or, $\tilde{p}=(0\,, \bar{n}_B\cdot \tilde{p} \,,\tilde{p}_\perp)$ and a residual momentum $k\sim \lambda^2 Q$ is the order of ultra-soft mode $\lambda^2  Q$. One can repeat the same power counting for the momenta in the jet region. 

Now momenta and fields can be expended as series sum of label momenta and fields which gives us basic building blocks of SCET.  
Neglecting the power correction $\mathcal{O}(\lambda^2)$, we match the current  $J^\mu(x)=\bar{\psi}\gamma^\mu \psi(x)$ in \eq{Wadef} onto the operators in SCET,
\bea
J^\mu(x) = \sum_{n_1, n_2}\int d^3\tilde p_1 d^3\tilde p_2 
  e^{i(\tilde p_1 - \tilde p_2)\cdot x} 
C_{\alpha\beta}^{\mu}(\tilde p_1,\tilde p_2)
  \cO^{\alpha\beta}(\tilde p_1,\tilde p_2;x) 
\,,
\label{eq:matching} \eea
where $n_1,n_2$ are unit four vectors in two directions and those are to be aligned with $n_B$ and $n_J$ during their sum and $\tilde p_{1,2}$ are associated label momenta $\tilde{p}_i=\frac{\omega_i n_i}{2} + \tilde{p}^\perp_i $. The matching coefficient $C$ carries spin indices $\alpha,\beta$. The SCET operator $\cO$ after BPS field redefinition \cite{Bauer:2001yt} is given by
\be
\cO^{\alpha\beta}(\tilde p_1,\tilde p_2;x) =
  \bar\chi_{n_1,\tilde p_1}^{\alpha j}(x) T[Y_{n_1}^\dag Y_{n_2}]^{jk}(x)
  \chi_{n_2,\tilde p_2}^{\beta k}(x) \,. 
\label{eq:opt_qq} \ee
Here we just write the quark part and drop the gluon part, which is not relevant in DIS \cite{Kang:2013nha}. 
The quark jet fields $\chi_{n,\tilde p}(x)$ are composed of $n$-collinear quark 
fields $\xi_n(x)$ and Wilson lines $W_n(x)$ expressed as 
\bea
\chi_{n,\tilde p}(x) = \big[ \delta(\omega - \bn\mcdot \cP)
  \delta^{(2)}(\tilde p_\perp - \cP_\perp)W_{n}^\dag \xi_{n} \big](x)
\,,
\label{eq:chi}
\eea
where $\cP^\mu$ is a operator \cite{Bauer:2001ct} that measures the label momentum of collinear fields $\cP^\mu\chi_{n,\tilde p} = \tilde p^\mu \chi_{n,\tilde p}$. When it is summed over all label momenta $\chi_n(x)=\sum_{\tilde p}\chi_{n,\tilde p} (x)$. The $n$-collinear Wilson line is  
\be
  W_{n}(x) = \sum_{\text{perms}}
   \exp\left[-\frac{g}{\bn\mcdot\cP}\bn\mcdot A_{n}(x)\right]\,,
\label{eq:Wn}\ee
where $A_n^\mu(x) = \sum_{\tilde p}A_{n,\tilde p}^\mu(x)$ is a $n$-collinear gluon field.

The soft gluon Wilson line can be written in the fundamental representation, for incoming states along ${n_i}=n_B$, as
\be
Y_{n_B}(x) = P\exp \left[ig\int_{-\infty}^0 d\xi \,n_B\cdot A_s(n_B \xi+x)\right]\,.
\label{eq:Ydef}\ee
In case of ${n_i}=n_J$ $i,e.,$ for outgoing states the path for $Y_{n_J}$, in \eq{Ydef}, turns into $0$ to $+\infty$ \cite{Arnesen:2005nk,Chay:2004zn}. 
  
After the BPS field redefinition 
\cite{Bauer:2001yt}, the measurement operator $\hat{\tau}_a$ in \eq{hatta} is also split up linearly into decoupled beam-collinear, jet-collinear and soft components each of them again constructed respective momentum flow operators of SCET
\be 
\hat{\tau}_a  = \hat{\tau}^{c_B}_a + \hat{\tau}^{c_J}_a + \hat{\tau}^{S}_a  
\,.
\label{eq:taua2}\ee
Henceforth we write the collinear operators only by $ \hat{\tau}^{J}_a,  \hat{\tau}^{B}_a$ dropping the $^c$ superscript. The soft part of the decomposition can be also split up into two operators $ \hat{\tau}^{S}_a= \hat{\tau}^{S_J}_a + \hat{\tau}^{S_B}_a$ depending on the jet and beam contributions.  
Similarly, the final state $X$ in \eq{hatta} is also decomposed into three sectors as $|X\rangle=|X_{cB}\rangle|X_{cJ}\rangle|X_S\rangle$ and each operator in \eq{taua2} matches with states in each sector.

After inserting the matching result \eq{matching} into \eq{Wadef}, the hadronic tensor has a factorized form. After having several simplifying steps as shown in \appx{interm}, we obtain
\begin{align}
W_{\mu \nu}(x,Q^2,\ta) = \bigg(\frac{8 \pi}{n_J\mcdot n_B}\bigg) & 
\int d\ta^J \, d\ta^B\, d\ta^S\, 
\delta\Bigl(\ta-\ta^J-\ta^B-\ta^S\Bigr) 
\nn\\
&\times H_{\mu\nu}(q^2,\mu) \cB_i(\ta^B,x,\mu) J(\ta^J,\mu) S(\ta^S,\mu) 
\label{eq:Wfact}\end{align}
To express the cross section in \eq{sigdef}, we contract the leptonic tensor with the hard function $H_{\mu\nu}$ in \eq{LH} and re-express it in terms of born-level cross-section and the scalar hard coefficient in \eqs{H}{born}. Now the differential cross-section can be written as
\begin{align}
\frac{d\sigma}{dx dQ^2 d\tau_a} = \frac{d\sigma_0}{dx dQ^2}& \int d\ta^J \, d\ta^B\, d\ta^S\,  
\delta\Bigl(\ta-\ta^J-\ta^B-\ta^S\Bigr)
\nn\\
&\times \sum_{i=q, \bar q} H_i(Q^2,\mu) \cB_i(\ta^B,x,\mu)J(\ta^J,\mu) S(\ta^S,\mu)
\label{eq:fact} 
\end{align}
The sum over $q$ goes over light quark flavors $q=\{u,d,s,c,b\}$ and the functions $\cB_q$, $J$, $S$ are quark beam, jet, and soft functions for angularity observables defined in \eqss{J_def}{Ses}{B_def}.
Jet and soft functions are computed analytically in \cite{Hornig:2009vb} at one-loop and numerically in \cite{Bell:2018vaa} for soft function and in \cite{Bell:2018gce} for jet function at two-loop. The angularity beam function is defined for the first time in this paper and we show one-loop result in \sec{beamfunc}.

\section{Resummation in Laplace space} \label{sec:resum}

The cross section in \eq{fact} contains logarithms of $\ta$, which shows singular behavior in small $\ta$ limit and spoils the convergence of perturbation theory and these logarithms can be resummed in the Fourier space \cite{Ligeti:2008ac,Abbate:2010xh} or, in the Laplace space \cite{Becher:2006mr,Becher:2006nr}. Here, we work in the Laplace space, where the transformation is defined as
\bea 
G(\nu, \mu) = \int^\infty_0 d\ta e^{-\nu \ta} G(\ta,\mu) 
\label{eq:LP}
\,,\eea
where $\nu$ is the variable in Laplace space conjugate to $\ta$.
After resummation, we come back to $\ta$ space by performing the inverse transformation : 
$(2\pi i)^{-1}\int^{\nu_0+i\infty}_{\nu_0-i\infty} d\nu \,\exp(\nu\ta) G(\nu)$
Because the resummation procedure is pretty standard, in the beginning of the section we make a quick summary of resummation procedure and for those who want to follow each step closely, we give more details in following subsections.

The cross section in \eq{fact} written as convolutions of beam, jet and soft functions in the momentum space is simply written as products of the functions in Laplace space denoted as $\tcB, \wt J, \wt S$ 
\bea 
\wt{\sigma}_q (\nu)
&=&
H_q(Q^2,\mu)\,  \tcB_q(\nu,\mu) \, \wt{J}(\nu,\mu) 
\wt{S}(\nu,\mu)  
\,,
\label{eq:tsig}\eea
where $\wt{\sigma}_q$ is simplified cross section and we should take the flavor sum $q$ and multiply by the born cross section $d\sigma_0/(dxdQ^2)$ to obtain \eq{fact} after the inverse transformation. The functions $\wt{J}, \tcB$ and $\wt{S}$ are the Laplace transformed jet, beam and soft functions. To avoid abuse of notation, we made dependencies on other variables than $\nu$ and $\mu$ implicit in Laplace space and will make them explicit if necessary.
The large logs in functions $G=\{H,\tcB,\wt{J},\wt{S}\}$ are resummed by the renormalization group (RG) evolution starting from natural scales $\mu_G$, where the logs are small, to the desired scale $\mu$. Renormalization group equations (RGE) of all functions in our factorization have the same structure as shown in \eqs{dGdmu}{gamG} and the solution is given by 
\be
G(\nu,\mu)=G(\nu,\mu_G)\, e^{K_G(\mu_G,\mu)+j_G \eta_G(\mu_G,\mu) L_G} 
\,,\label{eq:Gnu}\ee
where evolution kernels $K_G$ and $\eta_G$ are integration of the anomalous dimensions in \eq{Keta-def}, $j_G$ are constants in \eq{jkG}, the characteristic logarithm
$L_G$ contains $\nu$ in its argument as shown in \eq{LG} and the function $G(\nu,\mu_G)$ contains single and double log terms $L_G, L_G^2$ at order $\as$.
Inserting \eq{Gnu} into \eq{tsig} gives resummed cross section in Laplace space. 
To make inverse transformation easy, we replace the log terms $L_G$ in the function $G(\nu,\mu_G)$ in \eq{Gnu} by a derivative operator $\partial_{\eta_G}/j_G$ while we retain the log term on the exponent so that the derivative turns into the log when it hits the exponent. 
\bea
g(\partial_{\eta_G})= G(\nu,\mu_G)\left\vert_{L_G \to \partial_{\eta_G}/j_G}\right.
\,,
\label{eq:gdef}\eea
where $g=\{\tilde b, \tilde j, \tilde s\}$ are the functions in terms 
of the derivative operator.
This replacement makes these functions independent of variable $\nu$ hence unchanged under the inverse transformation. The evolution term $\exp[j_G\, \eta_G\, L_G]\sim \nu^{-\eta_G}$ only changes as in \eq{Linv}. After rewriting $\wt{\sigma}(\nu)$ as products of $\tilde b, \tilde j, \tilde s$ and collecting all $L_G$ terms from evolution factors in each function we have
\bea 
\tilde{\sigma}_q(\nu) &=& 
H_q(Q^2,\mu_H) ~
\tilde{b}_q(\partial_{\eta_B})~
\tilde{j}(\partial_{\eta_J})~ 
\tilde{s}(\partial_{\eta_S})   
\nn \\
&&\times  
e^{\kappa(\{\mu_i\},\mu)} ~\bigg( \frac{\mu_H}{Q}\bigg)^{-\eta_H(\mu_H,\mu)}
\bigg(\frac{\mu_J}{Q}  \bigg)^{-j_J\eta_J(\mu_J,\mu)}
\bigg(\frac{\mu_B}{Q}\bigg)^{-j_B\eta_B(\mu_B,\mu)} 
\bigg(\frac{\mu_S}{Q} \bigg)^{-j_S\eta_S(\mu_S, \mu)}
\nn\\
&& \times
\left(\nu e^{\gamma_E} \right)^{-\Omega}
\,, 
\label{eq:fact_LP2}\eea
where $\{\mu_i\}$ means $\mu_H,\mu_J,\mu_B,\mu_S$ and $\kappa$ and $\Omega$ are sums of the evolution kernels
\bea
\kappa(\{\mu_i\},\mu) &=& K_H(\mu_H,\mu)+K_J(\mu_J,\mu)+K_B(\mu_B,\mu)+ K_S(\mu_S,\mu)
\label{eq:kappa}\\
\Omega &=&  \eta_J(\mu_J,\mu)+\eta_B(\mu_B,\mu)+ \eta_S(\mu_S,\mu)
\,.
\label{eq:Omg}\eea
Taking the inverse Laplace transformation we can write the cross-section in the momentum space as
\bea 
\frac{d\hat{\sigma}_q}{d\ta}
&=& 
H_q(Q^2,\mu_H) ~
\tilde{b}_q(\partial_{\eta_B}) ~
\tilde{j}(\partial_{\eta_J})~ 
\tilde{s}(\partial_{\eta_S})
\nn\\
& &\times 
e^{\kappa(\{\mu_i\},\mu)}~ \bigg( \frac{Q}{\mu_H}\bigg)^{\eta_H(\mu_H,\mu)}  
\bigg(\frac{Q}{\mu_J}\bigg)^{j_J\eta_J(\mu_J,\mu)} 
\bigg(\frac{Q}{\mu_B}\bigg)^{j_B\eta_B(\mu_B,\mu)}
\bigg(\frac{Q}{\mu_S}\bigg)^{j_S\eta_S(\mu_S, \mu)}
\nn\\
&& \times 
\frac{\ta^{-1+\Omega}}{ \Gamma(\Omega)}e^{-\gamma_E \Omega}
\,.
\label{eq:sigta}
\eea
Up to including the flavor sum and the born cross section, \eq{sigta} is the resummed version of \eq{fact}.
A \emph{cumulative} cross-section is another conventional way to express the resummed results 
\be
\sigma_q^\text{cum}(\ta)
= \int_0^{\ta} d\ta^\prime\, \frac{ d\hat{\sigma}_q}{d\ta^\prime}
=\frac{ d\hat{\sigma}_q}{d\ta}\,\frac{\ta }{\Omega}
\,.
\label{eq:sigcum}
\ee
In rest of the section, we discuss more details about Laplace transformation and resummation. Readers familiar with technical step may skip following subsections.

\subsection{Renormalization group evolution} \label{sec:RGE}
In this subsection, we discuss RGE in the Laplace space and its solution that evolves functions of factorization from one to another scales and resum large logarithms.
 
The RGE of $\wt S, \wt J, \tcB$ in the Laplace space and also the hard function $H$ can be written as
\be\label{eq:dGdmu}
\mu\frac{d}{d\mu} G(\nu,\mu)= \gamma_G(\mu) \, G(\nu,\mu)
\,,\ee
where $G={H,\wt S, \wt J, \tcB}$ and $\gamma_G$ is their anomalous dimensions. Although the hard function $H$ is not transformed it is included in \eq{dGdmu} since its RGE structure is the same as other functions. The jet and beam functions are defined by the same collinear operator and their anomalous dimensions are the same $\gamma_{\wt J}(\mu)=\gamma_{\wt B}(\mu)$. We just give the anomalous dimension of the jet function and do not separately give that of the beam function.
The $\gamma_G$ takes following structure
\be\label{eq:gamG}
\gamma_G(\mu) =j_G \, \kappa_G \Gamma_\text{cusp} (\as) L_G +\gamma_G (\as)
\,,\ee
where $\Gamma_\text{cusp}(\as)$ and $\gamma_G (\as)$ are the cusp and non-cusp anomalous dimensions. The characteristic logarithm $L_G$ is defined as
\begin{align}\label{eq:LG}           
            \qquad L_G&=\begin{cases}\ln\left(\frac{Q}{\mu}\right) & \qquad G=H
            \,,\\
            \ln\left[\frac{Q}{\mu}(\nu e^{\gamma_E})^{-1/j_G} \right] &\qquad G=\{\wt S,\wt J, \tcB\}
            \,,
\end{cases}      
\end{align}
The consistency relation followed by scale independence of cross section $d\sigma(\mu)/d\mu=0$ is given by $\gamma_H(\mu)+\gamma_{\wt S}(\mu)+2 \gamma_{\wt J}(\mu)=0$, which is valid for any values of $Q,\mu,\nu$ in \eq{gamG} and it turns into three consistency relations
\bea
 j_H\, \kappa_H+j_S\, \kappa_S +2 j_J\, \kappa_J &=&0
\,,\nn\\
 \kappa_S +2  \kappa_J &=&0
\,, \nn\\
 \gamma_H(\as)+\gamma_S(\as)+2\gamma_J(\as)&=&0    
\,.
\label{eq:consis}\eea
The constants $j_G$ and $\kappa_G$ are given by
\begin{align}
j_G &= \left\{1,1,2-a\right\}\,,  
\nn\\
\kappa_G &=\left\{4,\frac{4}{1-a}, -\frac{2}{1-a}\right\}  
\,,
\qquad\qquad G=\{H, S, J\}
\label{eq:jkG}
\end{align}

The universal cusp anomalous dimension $\Gamma_\text{cusp} (\as)$ and
non-cusp anomalous dimension $\gamma_G (\as)$ are expressed in powers of $\as$ as
\be
\Gamma_\text{cusp} (\as) =\sum_{n=0} \Gamma_n\,  \left(\frac{\as}{4\pi}\right)^{n+1}
\,,\qquad\qquad
\gamma_G (\as)=\sum_{n=0} \gamma^G_n\,  \left(\frac{\as}{4\pi}\right)^{n+1}
\label{eq:Ggn}
\,,\ee
where $\Gamma_n$ are given in \appx{anom} and one-loop result for $\gamma^G_n$ are given in \cite{Hornig:2009vb}
\be
\gamma^G_0=\left\{-12 C_F\,,0\,, 6C_F \right\} \qquad G=\{H, S,  J\}
\,,
\label{eq:gamG0}
\ee
which again satisfies the consistency in \eq{consis} at the order $\as$.
The two-loop hard anomalous dimension is well known~\cite{Idilbi:2006dg, Becher:2006mr} and available up to three-loops~\cite{Moch:2005id}
\be
\gamma^H_1 = -2C_F\Bigl[ \Bigl( \frac{82}{9} - 52\zeta_3\Bigr) C_A + (3-4\pi^2+48\zeta_3)C_F + \Bigl(\frac{65}{9} + \pi^2\Bigr)\beta_0\Bigr] 
\,.
\label{eq:gh1}\ee
The two-loop soft anomalous dimension is computed in \cite{Bell:2018vaa} and a simplified expression for the angularity is given in \cite{Bell:2018gce}.
\bea
\gamma^{S}_1&=&\frac{2}{1-a}\left[ \left( -\frac{808}{27}+\frac{11}{9}\pi^2 +28\zeta_3 -\Delta \gamma^{C_A}\right)C_F C_A 
+\left( \frac{224}{27}-\frac{4\pi^2}{9}-\Delta\gamma^{n_f} \right) C_F T_F n_f \right]
\nn\,,\\
\Delta \gamma^{C_A} &=&\int_0^1 dx \int_0^1 dy
\frac{32x^2(1+xy+y^2)[x(1+y^2)+(x+y)(1+xy) ]}{y(1-x^2)(x+y)^2(1+xy)^2}
\ln\left[ \frac{(x^a+xy)(x+x^a y)}{x^a(1+xy)(x+y)} \right]
\nn\,,\\
\Delta\gamma^{n_f} &=&
\int_0^1 dx \int_0^1 dy
\frac{64x^2(1+y^2)}{(1-x^2)(x+y)^2(1+xy)^2}
\ln\left[ \frac{(x^a+xy)(x+x^a y)}{x^a(1+xy)(x+y)} \right]
\,,
\label{eq:gam1s}\eea
where $\Delta\gamma^{C_A,n_f}$ vanishes in the thrust limit ($a\to 0$) and $\gamma^{S}_1$ becomes that of thrust.
The one for jet function is given by the consistency $\gamma^{J}_1   =-\tfrac12 (\gamma^H_1+\gamma^{S}_1 )$.

The solution of RGE is given by \eq{Gnu} where the terms $K_G, \eta_G$ on the exponent are defined by integration of the anomalous dimension in \eq{gamG}
\be\label{eq:intgam}
\int_{\mu_G}^{\mu}  \frac{d\mu'}{\mu'}\gamma_G(\mu')
= j_G \, \kappa_G \int_{\mu_G}^{\mu} \frac{d\mu'}{\mu'}\Gamma_\text{cusp} (\as) \left[-\ln(\mu'/\mu_G)+L_G(\mu_G) \right]+\int_{\mu_G}^{\mu} \frac{d\mu'}{\mu'}\gamma_G (\as)
\,,\ee
where we split $L_G(\mu')$ in \eq{gamG} into two logs above and then define three integrals after replacing $\tfrac{d \mu'}{\mu'}$ by $d\as/\beta(\as)$
\bea
K_\Gamma (\mu_G,\mu)&=&\int^{\as(\mu)}_{\as(\mu_G)} \frac{d\as}{\beta(\as)}\Gamma_\cusp (\as )\, \int^{\as}_{\as(\mu_G)} \frac{d\as'}{\beta(\as')}
\,,\nn\\ 
\eta_\Gamma (\mu_G,\mu)&=&\int^{\as(\mu)}_{\as(\mu_G)} \frac{d\as}{\beta(\as)}\Gamma_\cusp (\as ) 
\,,\nn\\
K_{\gamma_G} (\mu_G,\mu) &=&\int^{\as(\mu)}_{\as(\mu_G)} \frac{d\as}{\beta(\as)}\gamma_G (\as ) 
\,.\label{eq:Keta-def} 
\eea
In \appx{anom} we give results of integration done order by order in $\as$. By collecting $K_{\Gamma}$ and $K_{\gamma_G}$ together as below 
\bea
K_G(\mu_G,\mu) &=& -j_G\kappa_G \, K_\Gamma (\mu_G,\mu) +K_{\gamma_G} (\mu_G,\mu)
\,,\nn\\
\eta_G (\mu_G,\mu) &=& \kappa_G \, \eta_\Gamma (\mu_G,\mu)
\,.\label{eq:ker}
\eea
Now \eq{intgam} is expressed in the form of exponent in \eq{Gnu}.

\subsection{Fixed-order coefficients}
In this subsection we write fixed-order structure and coefficients of functions in factorized cross section at $\cO(\as)$. 

Hard function in \cite{Kang:2013nha} and jet and soft functions in \cite{Hornig:2009vb} have the same logarithmic structures, which can easily be reconstructed by using RGE in \eq{dGdmu}. We can perturbatively solve the REG order by order in $\as$ and truncate terms higher-order than $\cO(\as)$ 
In other words, we take the scale $\mu_G$ in \eq{Gnu} such that logarithmic terms of $G(\nu,\mu_G)$ vanish and it just contains only constant at $\cO(\as)$. Then, we expand the right side of \eq{Gnu} including the exponent expressed with \eq{Keta} up to $\cO(\as)$.
Then, we obtain
\be\label{eq:Gfo}
G^\fo(L_G, \mu)= 1+\frac{\as(\mu)}{4\pi}\left[- j_G\kappa_G \frac{\Gamma_0}{2} L_G^2-\gamma^G_0 L_G +c^G_1 \right]\,,
\qquad\qquad G=\{ H,\wt{S},\wt{J} \}
\ee
The fixed-order beam function further factorized into short-distance coefficient and the PDF is given in \sec{beamfunc}.
The coefficients $j_G,\kappa_G,\gamma_0^G$ of logarithmic terms are given in \sec{RGE} and the one-loop constant $c_1^G$ are known from \cite{Bauer:2003di,Manohar:2003vb} for $H$ and \cite{Hornig:2009vb} for $\tJ$ and $\tS$.

\bea
c_1^H &=& C_F\left(-16 +\frac{\pi^2}{3}\right) 
\,,\nn\\
c_1^\tS &=&-C_F\frac{\pi^2}{1-a}
\,,\nn\\
c_1^\tJ &=& \frac{C_F}{2-a}\left( 14-13 a-\frac{\pi^2}{6}\frac{8-20a+ 9a^2}{1-a}- 4f(a)\right)
\,,
\label{eq:c1}\eea
where the jet function constant contains the integral $f(a)$
\be\label{eq:fa}
f(a)=\int^1_0 dx\, \frac{2-2x+x^2}{x}\ln\left[(1-x)^{1-a}+x^{1-a} \right]
\,.\ee 

Although two-loop constants \cite{Idilbi:2006dg,Becher:2006mr,Bell:2018gce,Bell:2018oqa} of hard, soft, and jet functions are known, we obtain the beam function at the one-loop level and at this moment we can achieve NNLL accuracy, for which all ingredients listed in  \tab{logaccuracy} in \appx{anom} are available.
Now we can obtain resummed functions by replacing $G(\nu,\mu_0)$ in \eq{Gnu} by its fixed-order expression \eq{Gfo} and the resummed cross section in the Laplace space as shown in \eq{tsig}. 

\subsection{Inverse transformation}\label{sec:inverse}
Let us discuss the inverse transformation to go back momentum space.
We first discuss the transformation of individual functions $G$ are essentially similar to that of the cross section. Then, We remark on the change in the case of cross section.

As discussed in the beginning of \sec{resum} we rewrite the fixed-order expression \eq{Gfo} in terms of derivative operators as $L_G\to \partial_{\eta_G}/j_G$
\be\label{eq:gas}
g(\partial_{\eta_G})=G^\fo(L_G,\mu)\vert_{L_G\to \partial_{\eta_G}/j_G}
=1+\frac{\as(\mu)}{4\pi}\left[ -\kappa_G \frac{\Gamma_0}{2j_G}\partial^2_{\eta_G} -\frac{\gamma^G_0}{j_G} \partial_{\eta_G} +c^G_1 \right]
\,,\ee
where $g=\tilde j, \tilde s$ represents jet and soft functions and analogous beam function $\tilde b$ is given in \eq{bas}. Then, RG evolved result equivalent to \eq{Gnu} is 
\be\label{eq:GLG}
G(L_G,\mu)=\tilde g(\partial_{\eta_G})\, e^{K_G(\mu_G,\mu)+j_G \eta_G(\mu_G,\mu) L_G} 
\,,\ee
where $\nu$ dependency only appears through $L_G$ in \eq{LG}. 
For a moment, we use shorten conventions $K_G$ and $\eta_G$ for $K_G(\mu_G,\mu)$ and $\eta_G(\mu_G,\mu)$ and recover at the end of this subsection.
With the following identity of inverse transformation 
\be\label{eq:Linv}
\cL^{-1} \left\{ \nu^{-\eta_G}  \right\}= \frac{\tau_a^{\eta_G-1}}{\Gamma(\eta_G)} 
\,,\ee
The functions in momentum space is given by
\be\label{eq:Gtaua}
G(\tau_a,\mu)
=\frac{e^{K_G}}{\tau_a}\,  g(\partial_{\eta_G})\frac{e^{j_G \eta_G L_G(\tau_a)}}{ \Gamma(\eta_G)}
\,,\ee
where the logarithm in $\ta$ is defined by
\be
L_G(\tau_a)=\ln\left[ \frac{Q}{\mu_G} \left(\ta e^{-\gamma_E}\right)^{1/j_G} \right] 
\,,
\qquad G=\{S,J\}
\,.
\label{eq:LGtau}\ee

The cross section is products of beam, jet, and soft functions expressed in a form of \eq{GLG} then, $\eta_G$ in \eq{Linv} is replaced by the sum $\Omega=\eta_S+2\eta_J$. Finally we obtain \eq{sigta}.

We can further move all the exponents in \eq{Gtaua} in front of $g(\partial_{\eta_G})$ by shifting the derivative by $L_G(\ta)$
\be\label{eq:Gtaua2}
G(\ta,\mu)
=\frac{e^{K_G+j_G \eta_G L_G(\ta)}}{\ta}\,  g\left(\partial_{\eta_G}+j_G L_G(\tau_a) \right)\frac{1}{\Gamma(\eta_G)}
\,,\ee
where scales entering the function $g$ and $L_G$ are $\mu_G$, while both $\mu$ and $\mu_G$ enter $K_G$ and $\eta_G$ .
The operator applied to the Gamma function turns into poly-logarithms
\begin{align}\label{eq:partial12}
\left[\partial_{\eta_G}+j_G L_G(\tau_a) \right]\frac{1}{\Gamma(\eta_G)} &=\left[-\psi(\eta_G)+j_G L_G(\tau_a) \right]\frac{1}{\Gamma(\eta_G)} 
\,,\nn\\
\left[\partial_{\eta_G}+j_G L_G(\tau_a) \right]^2\frac{1}{\Gamma(\eta_G)} &=\left\{\left[-\psi(\eta_G)+j_G L_G(\tau_a) \right]^2-\psi^{(1)}(\eta_G)\right\}\frac{1}{\Gamma(\eta_G)} 
\,,\end{align}
where $\psi(x)=\Gamma'(x)/\Gamma(x)$ and $\psi^{(1)}(x)=d \psi(x)/d x$.
Note that in cross section,  
$\Gamma(\eta_G)$ on left side of \eq{partial12} replaced by $\Gamma(\Omega)$ hence, all $\eta_G$ on right side of \eq{partial12} should be replaced accordingly.
We finally obtain the cross section as
\bea 
\frac{d\hat{\sigma}_q}{d\ta}
&=& 
H_q(Q^2,\mu_H) ~
e^{\kappa(\{\mu_i\},\mu)}~ 
\bigg( \frac{Q}{\mu_H}\bigg)^{\eta_H(\mu_H,\mu)}  
\bigg(\frac{Q}{\mu_J}\bigg)^{j_J\eta_J(\mu_J,\mu)} 
\bigg(\frac{Q}{\mu_B}\bigg)^{j_B\eta_B(\mu_B,\mu)}
\bigg(\frac{Q}{\mu_S}\bigg)^{j_S\eta_S(\mu_S, \mu)}
\nn\\
& &\times 
\ta^{-1+\Omega}e^{-\gamma_E \Omega}
\nn\\
&& \times
\bigg[\tilde{b}_q\bigg(\partial_{\Omega}+j_BL_B(\ta),\mu_B\bigg)~
\tilde{j}\bigg(\partial_{\Omega}+j_J L_J(\ta),\mu_J\bigg) 
\tilde{s}\bigg(\partial_{\Omega} +j_S L_S(\ta),\mu_S\bigg)  
 \bigg]
\frac{1}{ \Gamma(\Omega)}
 \,. 
\label{eq:sigta2}
\eea
\section{Angularity beam function} \label{sec:beamfunc}

In this section we review structure of beam function, summarize our result for angularity beam function at $\mathcal{O}(\as)$ and show numerical results at different values of $a$. 
We concentrate to the region $a< 1$ as we are restricted in \SCETa region. We perform the details of computation in \appx{oneloop} and those who likes to know details may read the appendix.
 
The angularity beam function is the collinear matrix elements with initial proton state and it measures angularity $\ta$ as well as momentum fraction $x$ of off-shell quark having hard scattering.
If the scale of beam function $Q\ta^{1/(2-a)}$ is much larger than $\Lqcd$, it is factorized and matched onto proton PDFs $f_j(x)$ at the scale $\Lqcd$ with the matching coefficient $\cI_{ij}(\ta,x)$ as
\be\label{eq:BIf}
\cB_{q}(\ta,x)=\sum_{j=q,g}\, \cI_{qj}(\ta,x)\otimes f_{j}(x)
\,,\ee
where the symbol $\otimes$ represents convolution integral defined 
as $a\otimes b=\int_x^1 d\xi/\xi\, a(x/\xi)\, b(\xi)$. 
We do not compute the gluon beam function $B_{g/P}(\ta,x)$ which is not relevant in the process. 
In the Laplace space it has similar structure
\be\label{eq:tBIf}
\tcB_{q}(\nu,x)=\sum_{j=q,g}  \tcI_{qj}(\nu,x)\otimes f_{j}(x)
\,.\ee
Here, we give the expression of matching coefficient in Laplace space and the expression in momentum space is given in \appx{oneloop}. To the order $\as$, we have
\be\label{eq:In}
 \tcI _{qj}(\nu,z)= \mathbbm{1}_{qj} + \tcI^{(1)}_{qj}(\nu,z)
\,,\ee
where leading-order coefficients $\mathbbm{1}_{qj}= \delta_{qj} \,\delta(1-z)$ and $\tcI^{(1)}_{qj}$ represents the one-loop correction. Structure of the matching coefficient can be obtained by solving RGE in \eq{dGdmu} at fixed order in $\as$ as done in \eq{Gfo}.
But unlike soft and jet functions, one has to take the factorized form in \eq{tBIf} into account when solving the RGE in \eq{dGdmu}. Then, we obtain
\be\label{eq:tI1}
 \tcI^{(1)}_{qj}= \frac{\as}{4\pi}\left[\left( { -}j_B\kappa_B \frac{\Gamma_0}{2} L_B^2{ -}\gamma^B_0 L_B \right) \mathbbm{1}_{qj}+4 C_{qj} P_{qj}(z) L_B +\wtc_1^{qj}(z,a)
 \right]
\,, \ee
where $C_{qj}=C_F, T_F$ for $j=q,g$. One of the logarithmic terms $L_B$ is associated with PDF with the splitting functions $P_{qj}$ 
\bea
\label{eq:PqqPqg}
P_{qq}(z) &=& \left[\frac{\theta(1-z)}{1-z}\right]_+ (1+z^2) + \frac{3}{2}\delta(1-z) = \left[ \theta(1-z)\frac{1+z^2}{1-z}\right]_+ 
\,,\nn\\
P_{qg}(z) &=& \theta(1-z) [(1-z)^2 + z^2]\,.
\eea
In \eq{tI1}, the logarithmic terms are known since $j_B=j_J$, $\kappa_B=\kappa_J$ and $\gamma^B_0=\gamma^J_0$ given \eqs{jkG}{gamG0} whereas the constant $c_1^{qj}(z)$ is new result and not given in any literature to our best knowledge. 
The beam function in terms of the derivative operator similar to \eq{gas} is given by
\be
b_q(\partial_{\eta_G})=\sum_{j=q,g}  \left\{
\mathbbm{1}_{qj}+\frac{\as}{4\pi}\left[ \mathbbm{1}_{qj} \left( -\kappa_B \frac{\Gamma_0}{2 j_B} \partial^2_{\eta_B}-\frac{\gamma^B_0}{j_B}  \partial_{\eta_B} \right) +\frac{4 C_{qj} P_{qj}(z)}{j_B}  \partial_{\eta_B} +\wtc_1^{qj}(z,a)
\right]
\right\}\otimes f_j
\,,
\label{eq:bas}\ee
where the constant term $\wtc^{qj}_1$ are given by
\bea
 \wtc_1^{qq}(z,a) &=&2C_F\left[ \fraca (1+z^2) \cL_{1}(1-z) +\frac{\pi^2}{12}\frac{a(4-a)}{(2-a)(1-a)}\cL_{-1}(1-z) \right.
\nn \\ 		&&\qquad\left. + 1-z -\fraca\frac{1+z^2}{1-z}\ln z\right]
\,,  
\label{eq:wtc1qq}
\\
\wt{c}_1^{qg}(z,a) &=& 2T_F \left[1-P_{qg}(z) +\fraca P_{qg}(z)\ln\left( \frac{1-z}{z} \right)\right]
\,,
\label{eq:wtc1qg} \eea
where the distribution function $\cL_n(1-z)$ is defined as
\begin{align}\label{eq:cLn}           
            \cL_n(1-z)&=
            \begin{cases}
            \left[\frac{\ln^n(1-z)}{1-z}\right]_+ & \quad n\ge 0\,, \\
            \delta(1-z) &\quad  n=-1.
 \end{cases}      
\end{align}
The constants $\wtc^{qq}_1,\wtc^{qg}_1$ with $a=0$ reduces to those in thrust beam function \cite{Stewart:2010qs}.
Note that our factorization is valid in $a<1$ region. In the region, they are well-defined and smooth functions of $a$. 
Their momentum-space expressions are given at the end of \appx{oneloop}.

\begin{figure}[h]
\begin{center}
\includegraphics[scale=.32]{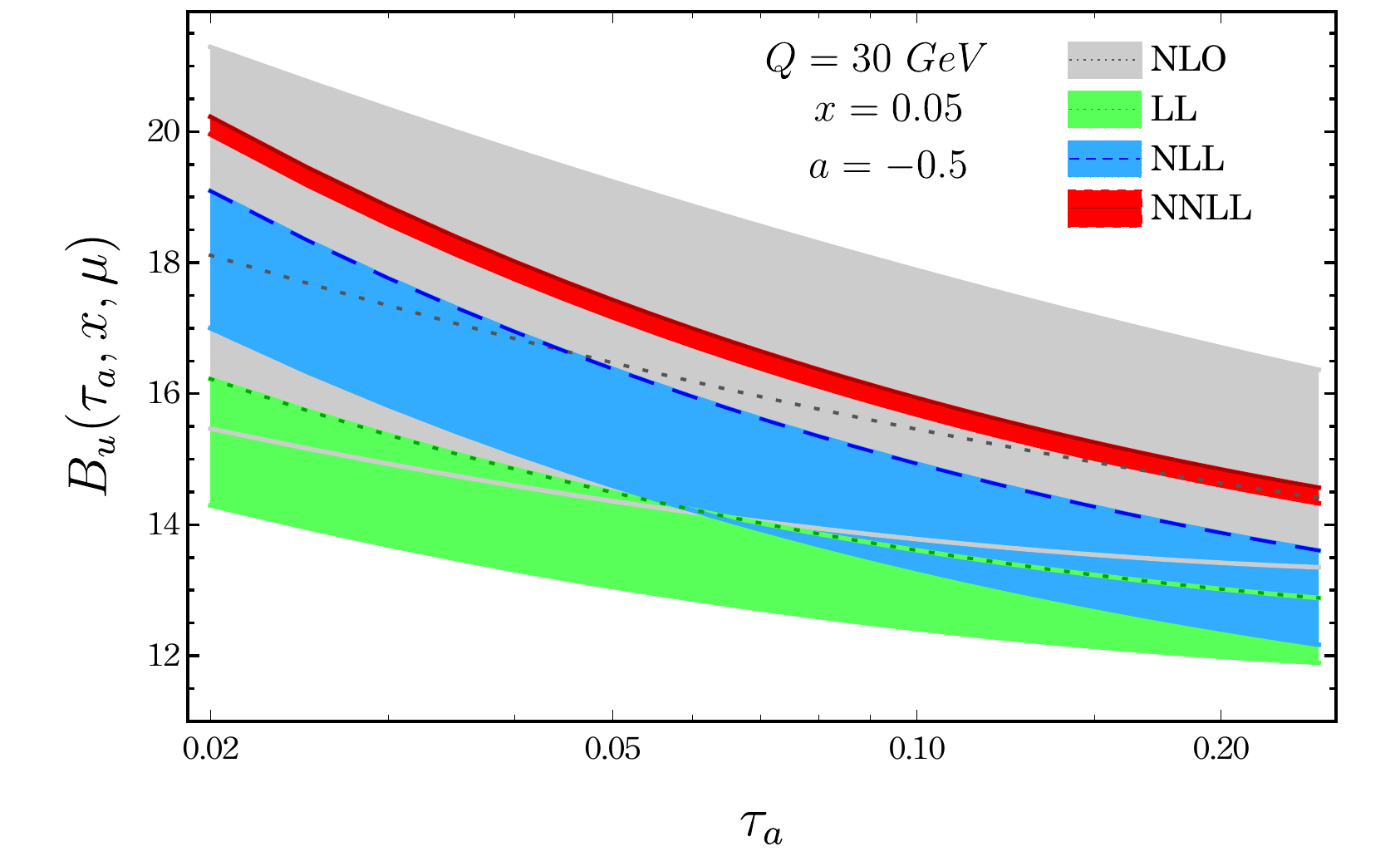}
\includegraphics[scale=.34]{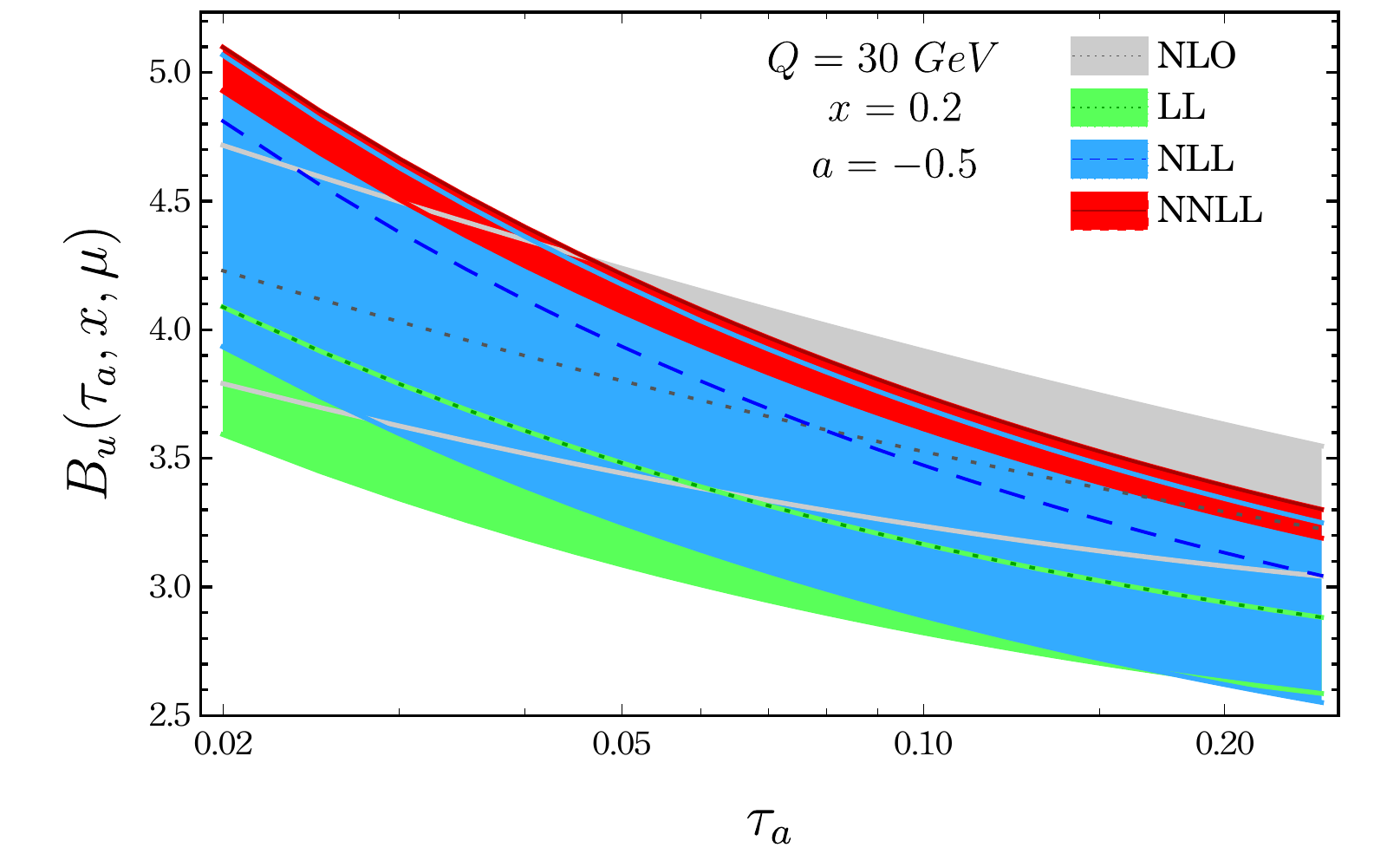} 
\includegraphics[scale=.33]{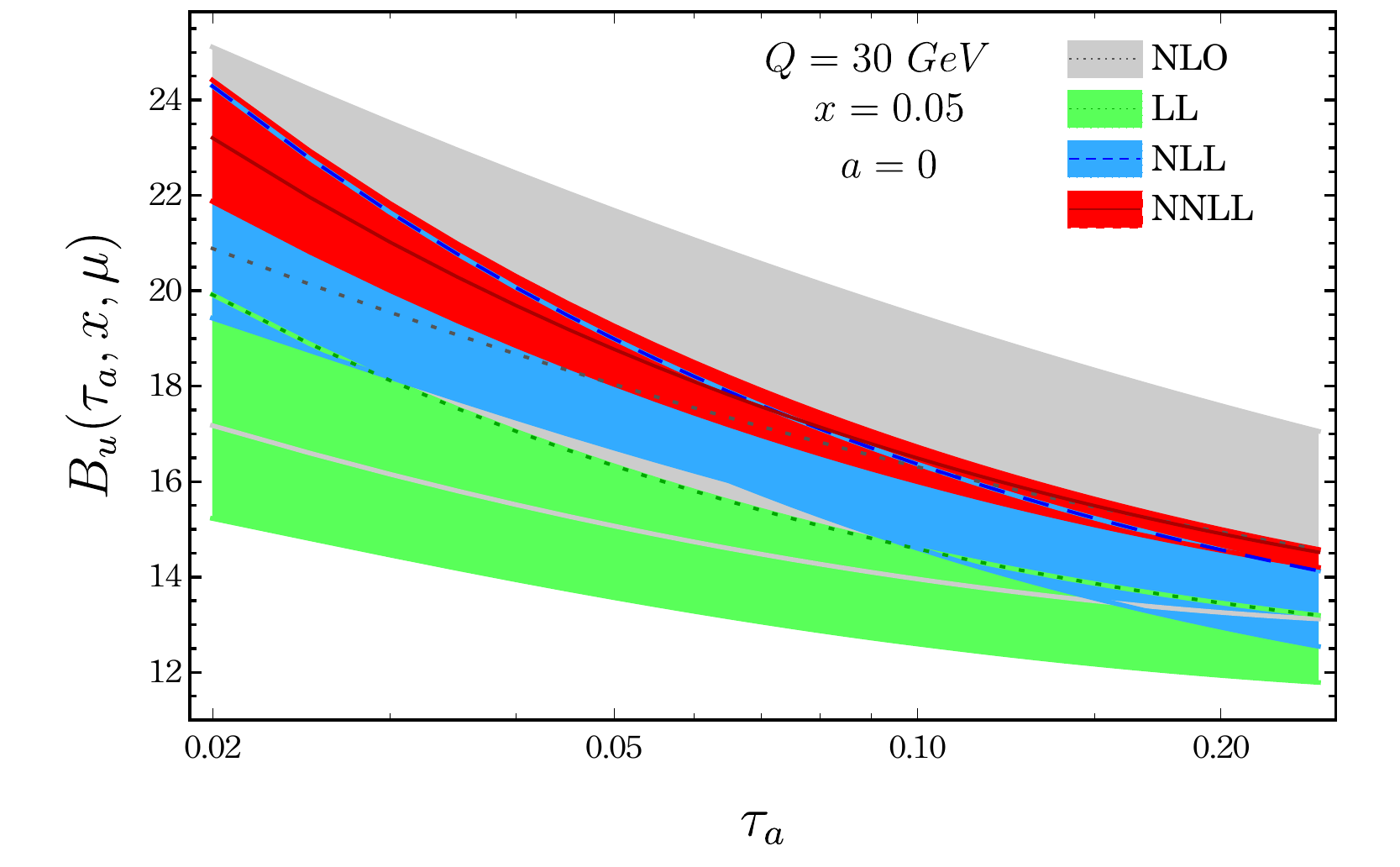} 
\includegraphics[scale=.33]{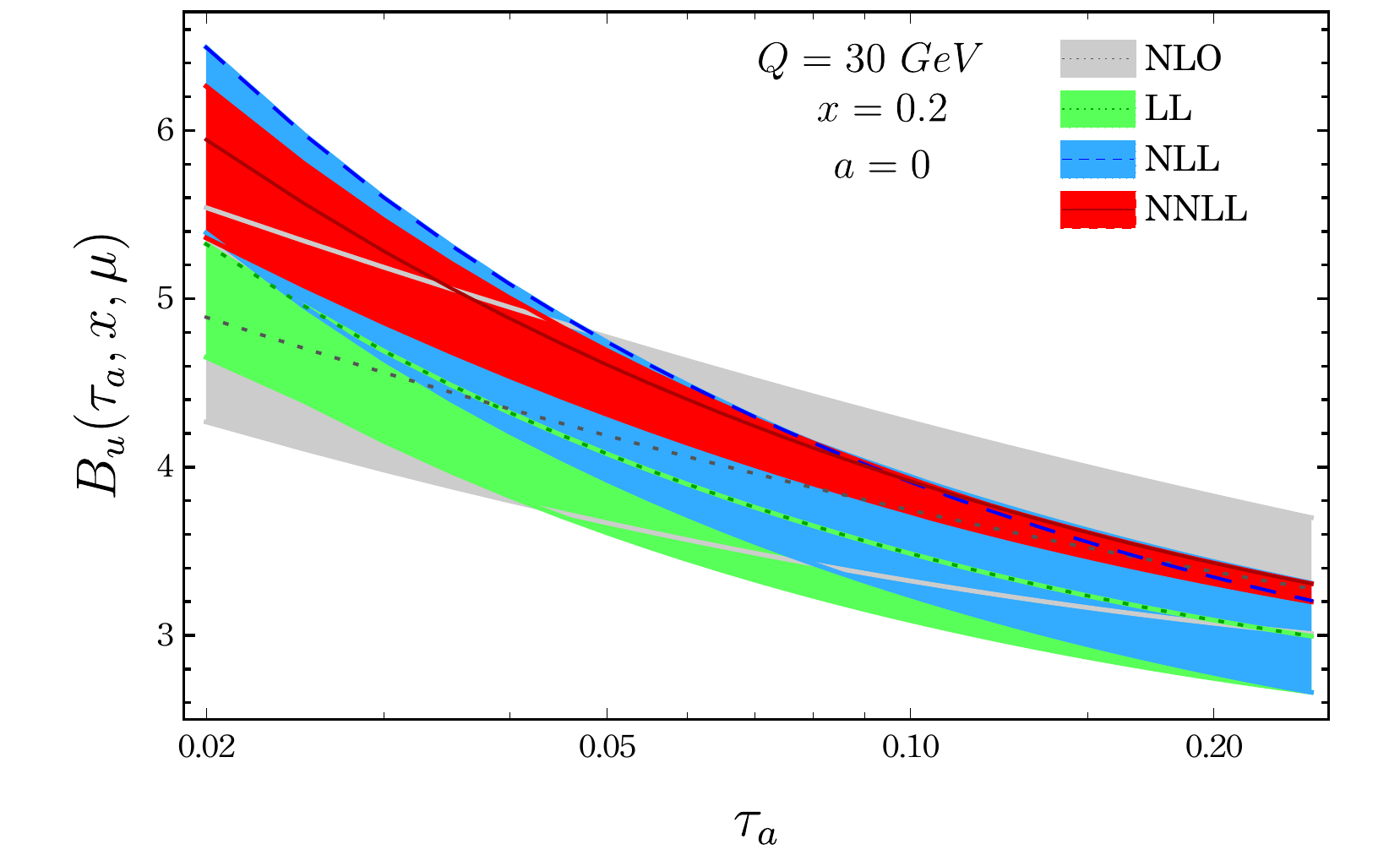} 
\includegraphics[scale=.33]{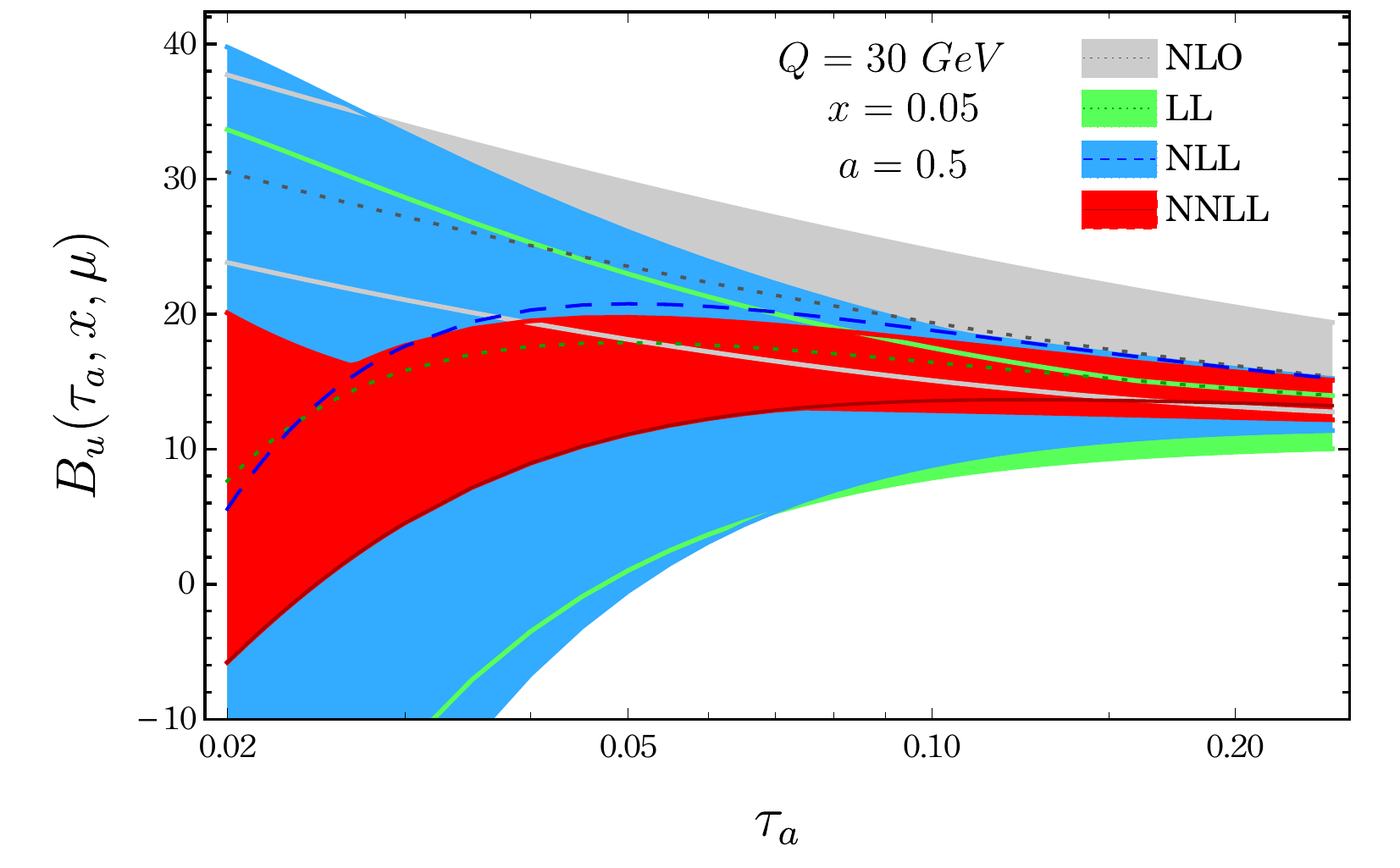}
\includegraphics[scale=.33]{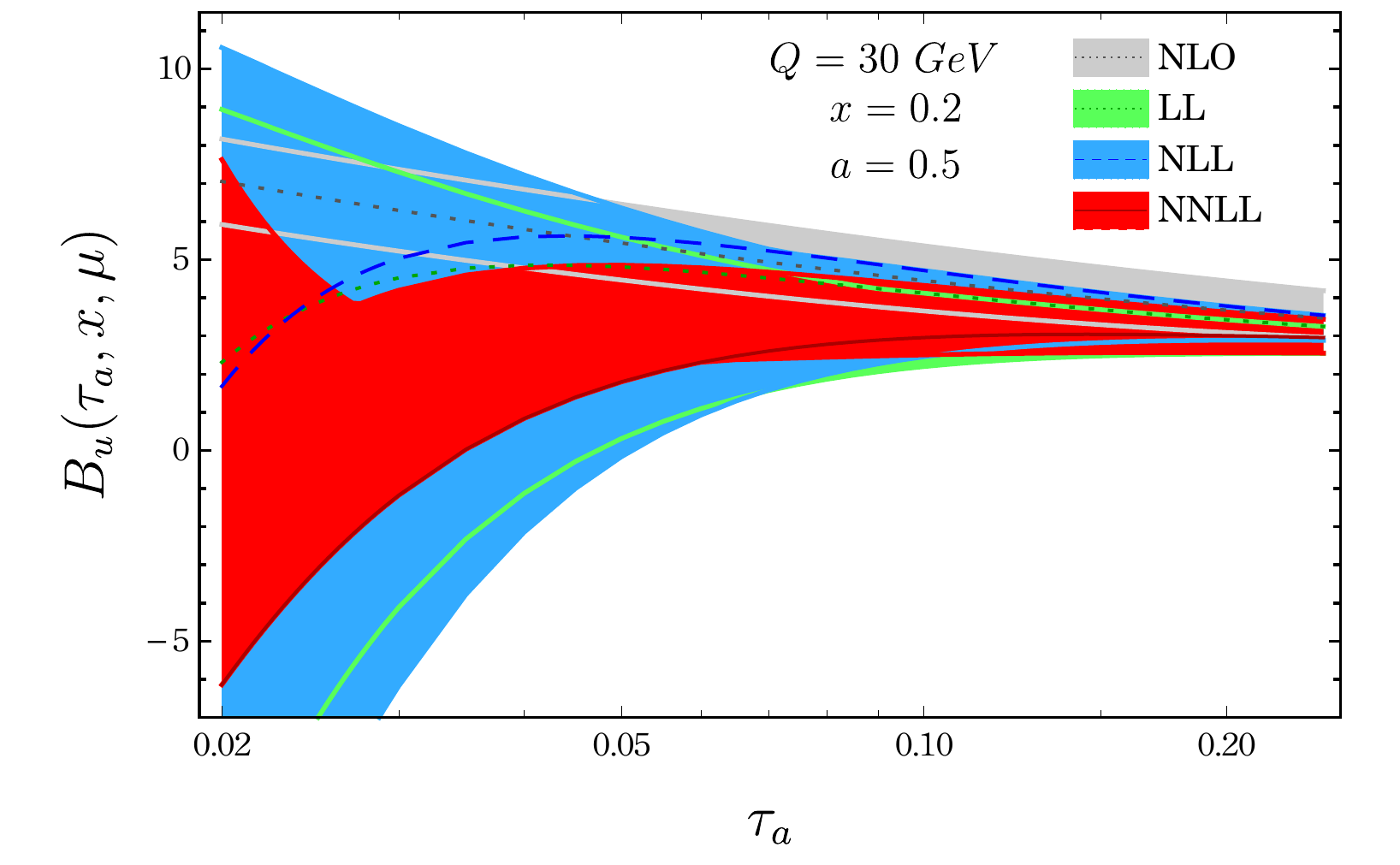}
\end{center}
\caption{ $\ta$ variation of beam function $B_u(\tau_a,x,\mu)$  for $u$ quark at $Q=30\GeV$.  Two columns are for  $x=0.05, 0.2$ and three rows are for $a=-0.5, 0, 0.5$ respectively. 
\label{fig:bf_tau}}
\end{figure}

Now we present the numerical results of our angularity  beam function. We compare the cumulative beam function
\be
B_q(\ta,x,\mu)
=\int_0^{\ta} d\ta' \,\cB_q(\ta',x,\mu) \label{eq:cB}
\ee
that captures delta function $\delta(\ta)$.

In \fig{bf_tau}
we show resummed results at LL, NLL, and NNLL accuracy as well as fixed-order result at NLO\footnote{Here, by NLO we mean fixed-order results at $\mathcal{O}(\alpha_s)$ from the factorized cross section in \eq{fact}, which captures all singular terms of full QCD cross section. The full QCD result is not included in this work and will be considered in future work.}.
For resummed result we evolve the function from canonical scale $\mu_B=Q\ta^{1/(2-a)}$ to a hard scale $\mu=Q$ and 
for the fixed-order we take $\mu=Q$. To estimate perturbative uncertainties
the scale $\mu_B$ is varied by a factor of two up and down
for resummed result while the scale $\mu$ is held fixed. In the fixed-order the scale $\mu$ is varied by factors of two. We use MMHT2014 PDF data set \cite{Harland-Lang:2014zoa} for the plot. The results are shown for $u$ quark at $Q=30 \GeV$. The rows are for different values of  angularity parameter $a=-0.5, 0$, and $0.5$ and the two columns are corresponding to the different $x=0.05, 0.2$. The numerical results show, the shape of $\ta$ distribution is sensitive to $a$ while less sensitive to $x$. The second row indicates the thrust results ($a=0$ limit) for angularity beam function. 
Perturbative uncertainties tends to smaller at $x=0.05$ due to typical pitching behavior of PDF around this value. 

At NLO, the beam function is related to fragmenting jet function (FJF) \cite{Procura:2009vm,Jain:2011xz,Ritzmann:2014mka} via crossing symmetry at the amplitude level and both are factorized in similar way in terms of perturbative matching coefficients and non-perturbative (NP) matrix elements: PDF and fragmentation functions. Therefore, their matching coefficients $\cI_{ij}(\ta,z)$ and $\cJ_{ij}(\ta,z)$ share many terms in common. It would be interesting to study a quantitative comparison coefficients between beam function and FJF. 

\comment{
\begin{figure}[t]
\begin{center}
\includegraphics[scale=0.45]{plots_DISang/Rq_ta1.pdf} 
\includegraphics[scale=0.45]{plots_DISang/Rq_ta25.pdf} 
\end{center}
\caption{ \label{fig:BfFF} 
 Variation of $R_q$ with $a$ for the $u$ quark. Left and right sub-figures are for two values of $\ta=0.1, 0.25$ respectively. NLO results are shown in black dashed lines and NNLL results are in orange lines.}
\end{figure}
}

The FJF includes the radiation around the measured fragmenting hadron at the final state. At first order in $\as$ in fragmenting jet an off-shell quark splits into two partons as $i^* \to k j $ whereas, in case of beam function an on-shell particle provides radiation $i \to k^* j$. One can obtain one from the other by using following crossing relation between splitting functions \cite{Ritzmann:2014mka}
\be 
P_{i\to k^*j}(2p_i\cdot p_j, x)=(-1)^{\Delta_f} P_{k^*\to ij}(-2p_i\cdot p_j, 1/x)
\,,
\label{eq:PP}\ee
where, $p_i,p_j$ are the collinear momentum and $\Delta_f$ is the difference in the number of incoming fermions. 
Squared matrix element for both functions remain the same under the crossing and the difference in 2-body collinear phase-space leads to a difference between them. 
Comparing our results \eqs{wtc1qq}{wtc1qg} with the angularity FJF Eqs.~(B.5)(B.15) in \cite{Bain:2016clc} after Laplace transformation one can notice that only the $\ln z $ term of \eqs{wtc1qq}{wtc1qg} and corresponding term differ and the differences are given by 

\bea
\Delta_{qj}(z,a) 
&=&\tcI^{(1)}_{qj}(\nu,z,a) - \wt{\cJ}^{(1)}_{qj}(\nu,z,a) 
\nn\\
&=&\frac{\as C_{qj}}{\pi}\frac{P_{qj}(z)}{2-a}
\ln \left[
\frac{z^{1-a} + (1-z)^{1-a} }{z^{2(1-a)}}
\right]
\label{eq:Dqj}
\,,\eea 
where $C_{qj}=C_F,T_F$ for $j=q,g$ and the splitting function $P_{qj}(z)$ is given in \eq{PqqPqg}. 
All the other terms in the beam function and FJF are cancelled out. 
Note that, the difference between the coefficients in the momentum space is the same as $\Delta_{qj}(z,a)$ up to multiplication of $\delta(\ta)$. 
 The differences is sensitive to $a$ and $z$ only and remains constant with respect to $\ta$. 
 
\begin{figure}[htbp]
\begin{center}
\includegraphics[scale=0.3]{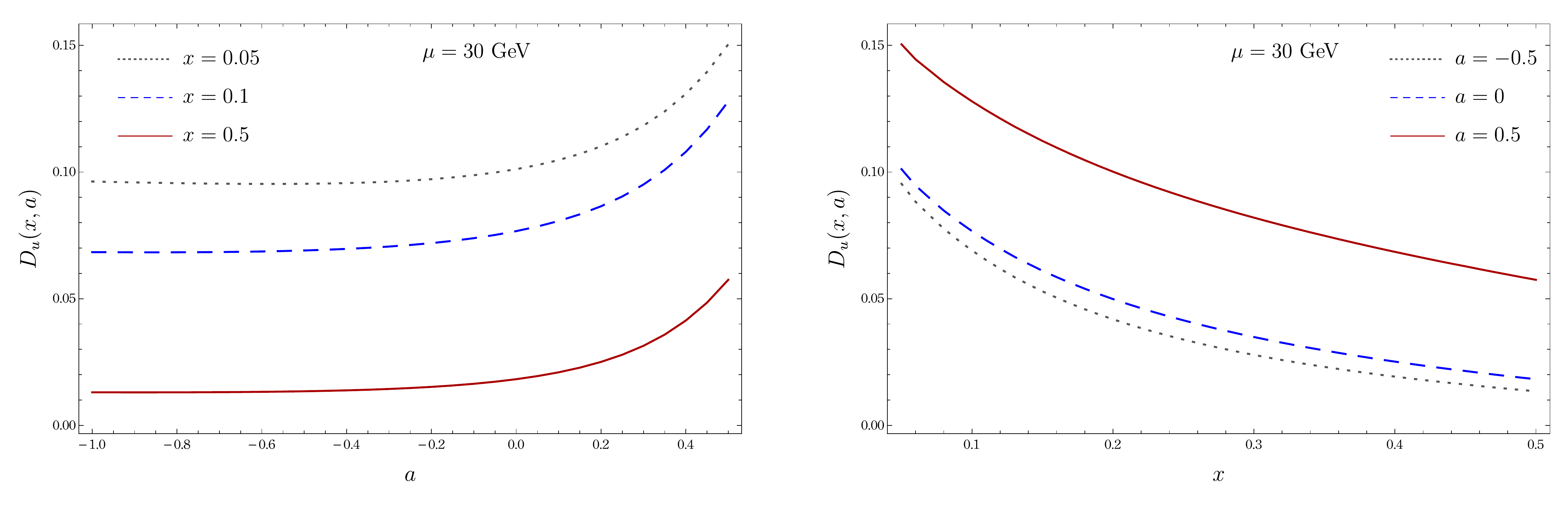} 
\end{center}
\caption{ \label{fig:Dq} 
The difference $D_u(x,a)$ between angularity beam function and FJF for the case of $u$ quark at $\mu=30\GeV$. The left and right sub-figures represent $a$ and $x$ dependency respectively. }
\end{figure}

To understand the impact in the beam function we define a relative change $D_{q}$ normalized by the leading-order beam function, which is simply the quark PDF
\be 
D_q(x,a) = \frac{1}{f_q(x,\mu)}\bigg[\sum_{j=q,g}\Delta_{qj}(x,a) \otimes f_j(x,\mu)\bigg]
\,.
\label{eq:Dq}\ee
\fig{Dq} shows the numerical results of $D_q$ for $u$ quark at $\mu=30\GeV$ in $a$ space and in $x$ space. 
It decreases monotonically as $a$ decreases and
also it slowly converges to a constant at large negative value of $a$ far below the region of $a$ in the plot. In positive region of $a$, the change is relatively faster because near $z=1$ \eq{Dqj} scales like $\Delta_{qq}\propto 1-a +(1-z)^{-a}$ and a dominant term among two terms is switched near $a=0$. Because the term comes from FJF, it is the feature of FJF and the beam function behaves rather slow.
On the other hand, increasing behavior with decreasing $x$ in \fig{Dq} are observed in both beam function and FJF in common.

\section{Angularity distributions at NNLL} \label{sec:numerical}
We now present our prediction for the angularity distribution in \eq{fact}, which resums large logarithmic terms to NNLL accuracy by using formula given in \eqs{sigta}{sigta2} and $\as$ coefficients of hard, jet, soft, and beam functions in \sec{resum} and \sec{beamfunc}. 
We are interested for the angularity parameter space well below $a=1$ and do not go to the further $a\to 1$ where power correction to the $\SCETa$ factorization cannot be neglected for such value of $a$. Ref.~\cite{Budhraja:2019mcz} discusses accessibility of this region of $a$, using recoil-sensitive factorization which is beyond scope of the work. 
Note that the cross section is differential in $x$ and $Q^2$ as well as in $\ta$ and we will show the dimensionless cross section normalized by the born cross section in \eq{born}
\be
\frac{d\hat{\sigma}}{d \ta} =\sum_{q}\frac{d\hat{\sigma}_q}{d \ta}
=\left(\frac{d\sigma_0}{dx dQ^2}\right)^{-1}\frac{d\sigma}{d \ta dx dQ^2}
\,,
\label{eq:hatsig}\ee
where $d\hat{\sigma}_q/d\ta$ is given by \eqs{sigta}{sigta2}.
As the differential cross-section falls rapidly in small $\ta$ region, we also multiply by a weight factor $\ta$ for better visibility. We choose $\sqrt{s} = 140\GeV$ as EIC plans to achieve $\sqrt{s}=45$ and $140~ \GeV$ \cite{AbdulKhalek:2021gbh,Accardi:2012qut}.
The coupling constant is taken as $\alpha_s(m_z)=0.118$ to have consistency to the central value of the MMHT2014 PDF data set \cite{Harland-Lang:2014zoa}. Throughout this paper we use the PDF data set MMHT2014 at NNLO and include five quarks and antiquark flavors and three-loop beta functions 
are used for consistency. Alternative choices such as NNPDF31 \cite{Bertone:2017bme}, CT18 \cite{Hou:2019qau}, HERAPDF20 \cite{Abramowicz:2015mha}, ABMP16 \cite{Alekhin:2017kpj} show the similar result and their differences are much smaller than our perturbative uncertainties at NNLL accuracy. As shown in the analytical form \eq{sigta}, the angularity differential-cross section is sensitive to the parameters longitudinal momentum fraction $x$, momentum transferred in the process $Q^2$ and  the angularity parameter $a$.  The sensitivity to the distribution on these parameters $x, Q^2$ and $a$ are studied numerically. 

\begin{figure}[htb]
\begin{center}
\includegraphics[scale=.45]{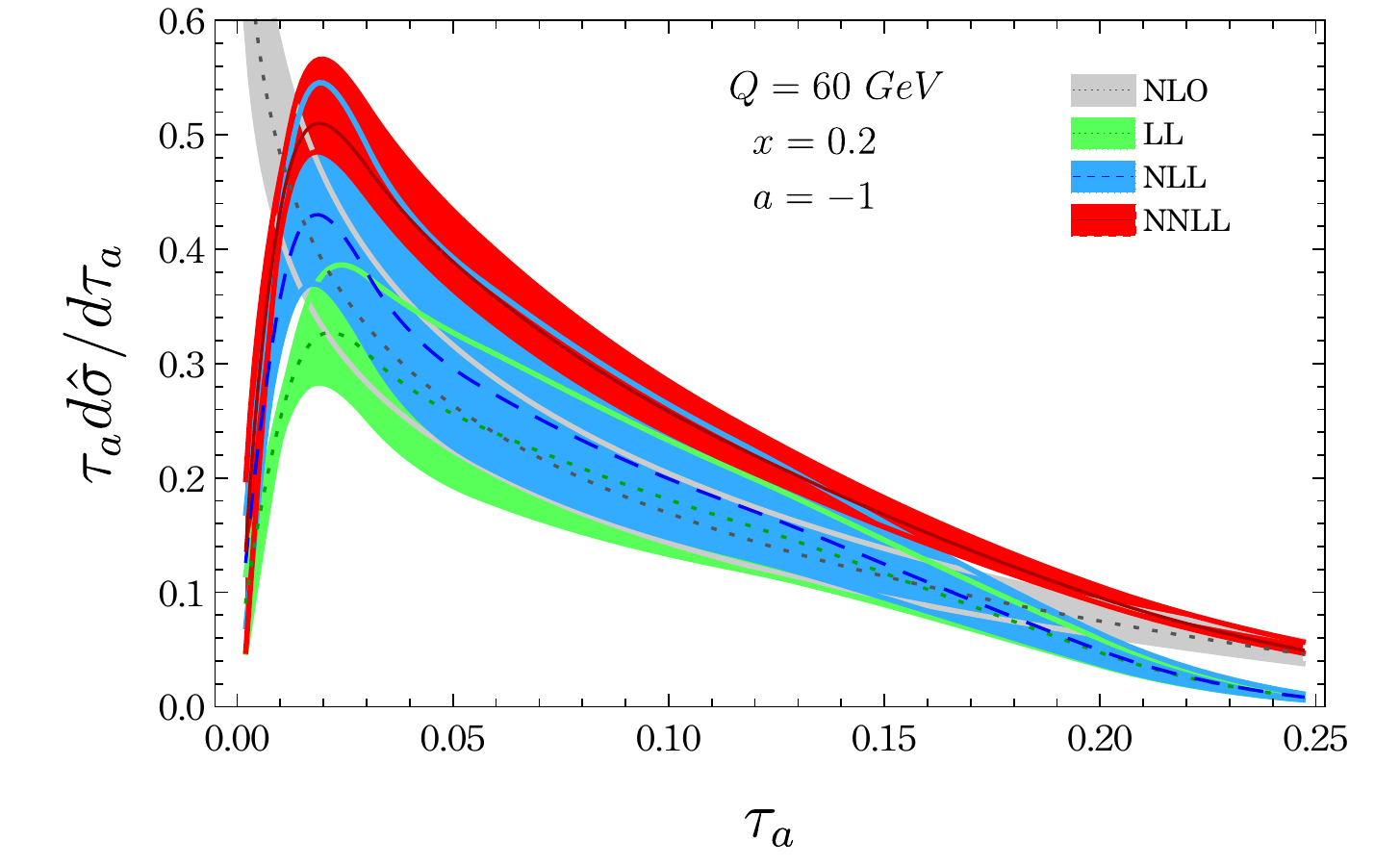} 
\includegraphics[scale=.45]{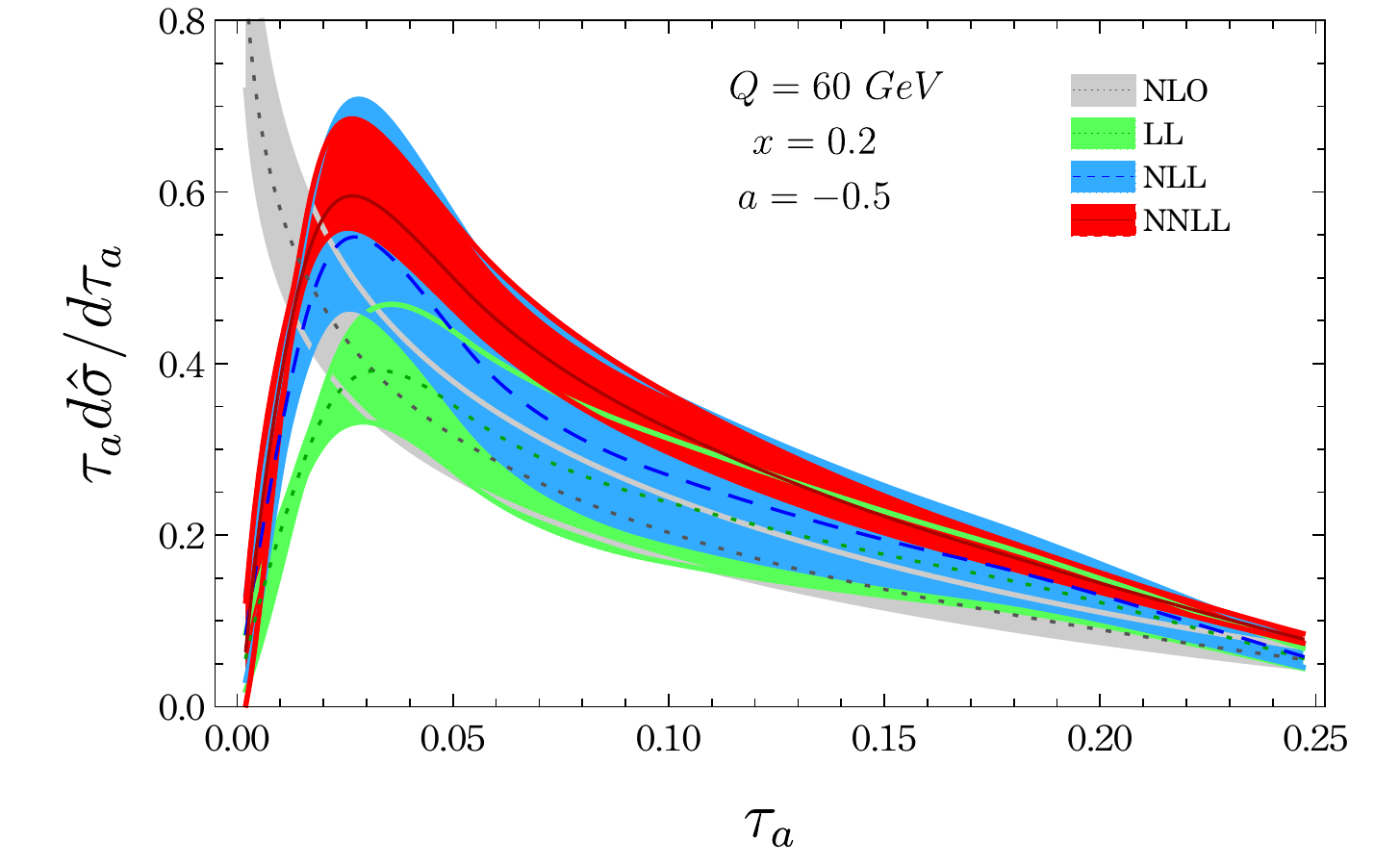} 
\includegraphics[scale=.45]{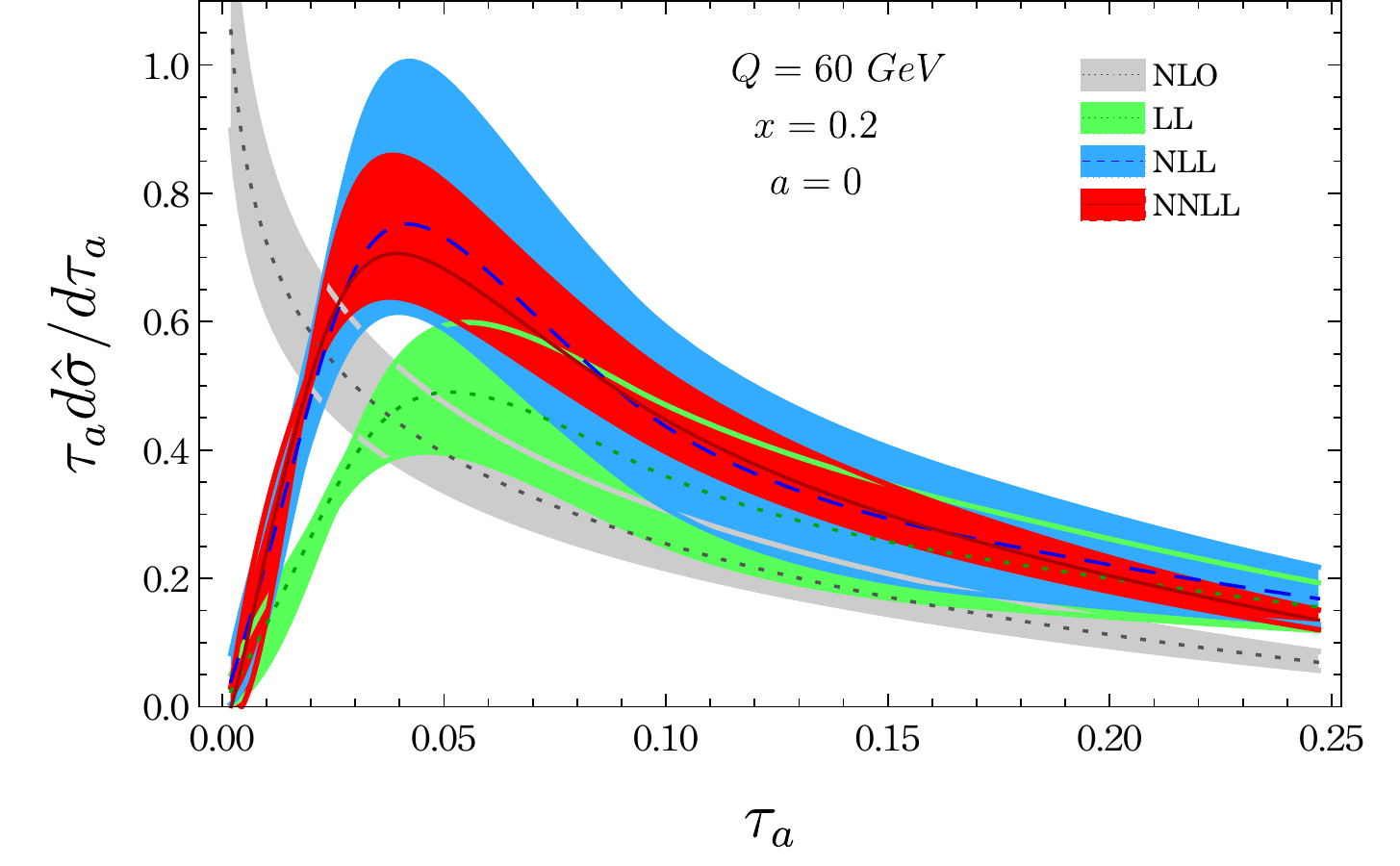} 
\includegraphics[scale=.45]{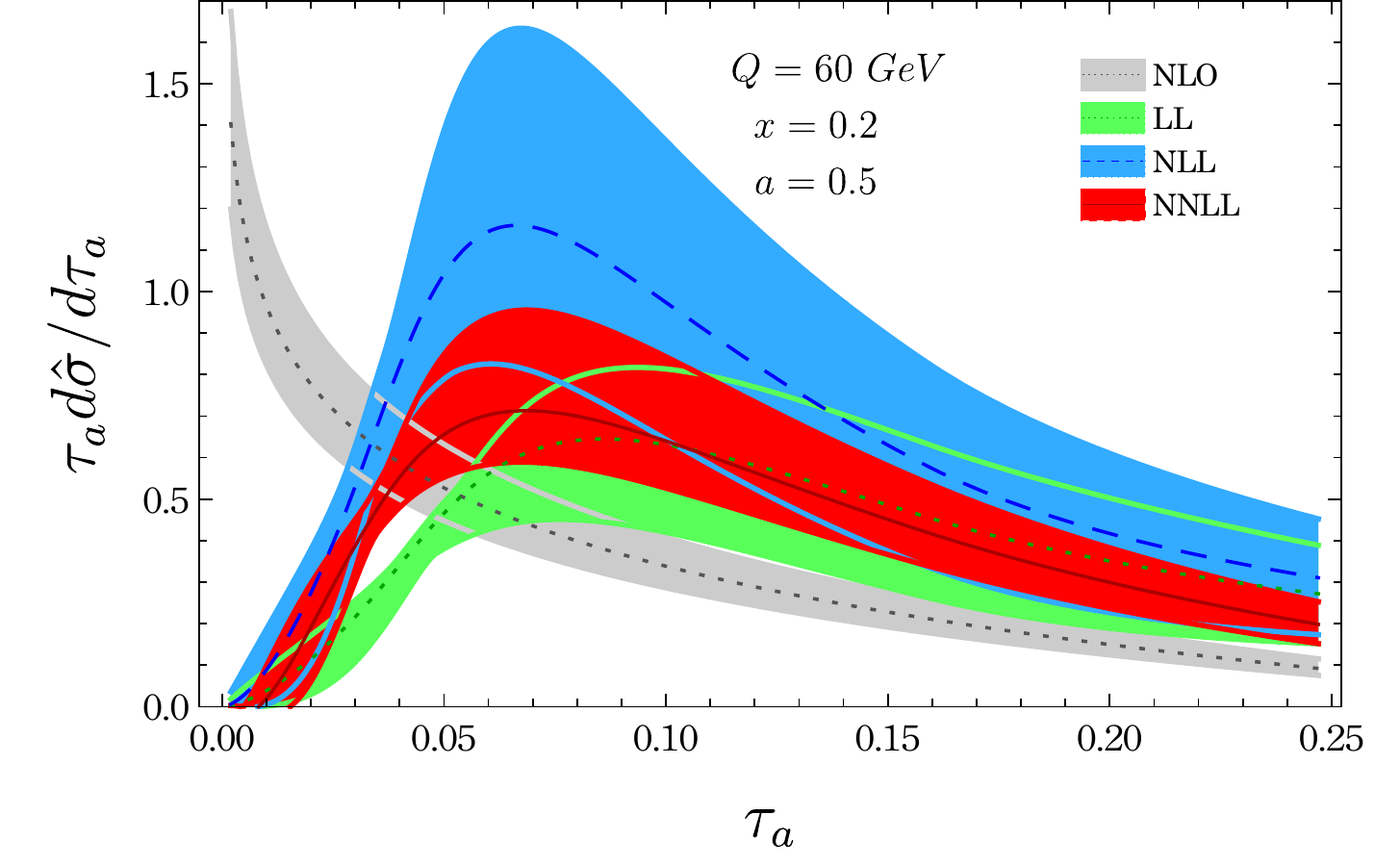} 
\end{center} 
\caption{Differential cross-section of DIS angularity $\ta$ for different $a=-1,-0.5,0,$ and $0.5$ at a fixed $x=0.2$ ($\sqrt{s}=140$ GeV). The bands indicate perturbative uncertainties. } \label{fig:dcsa}
\end{figure}
\begin{figure}[htb]
\begin{center}
\hspace{-1cm}\includegraphics[width=8.cm,height=3.2cm]{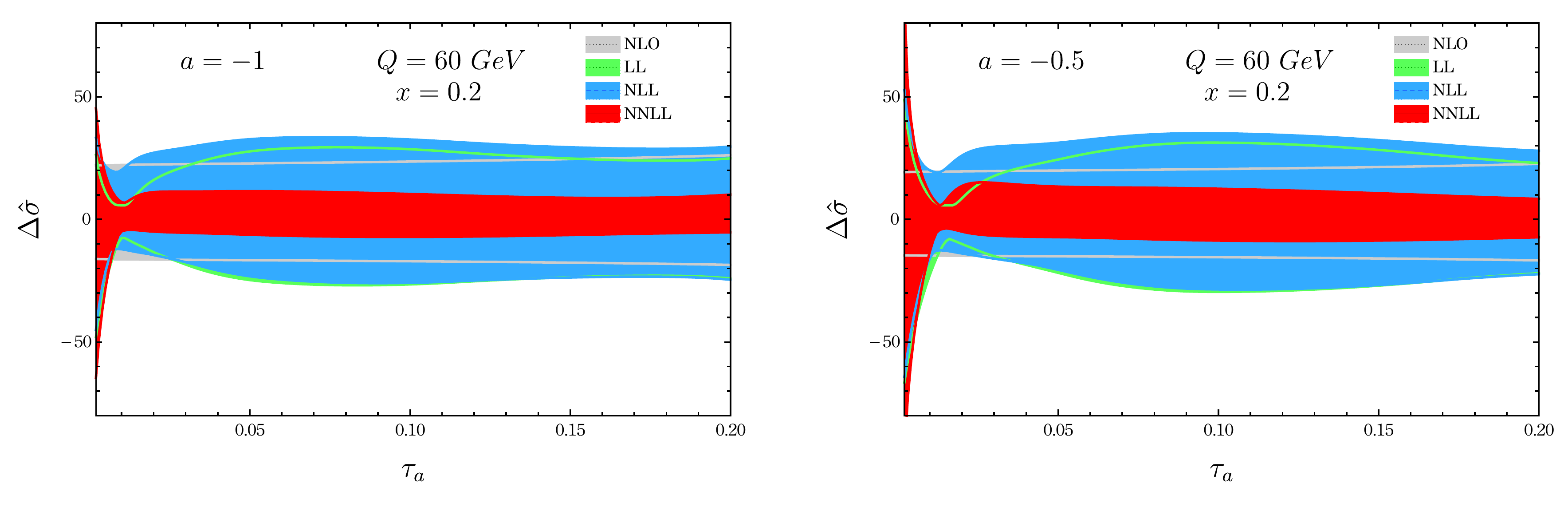}  
\includegraphics[width=8.cm,height=3.2cm]{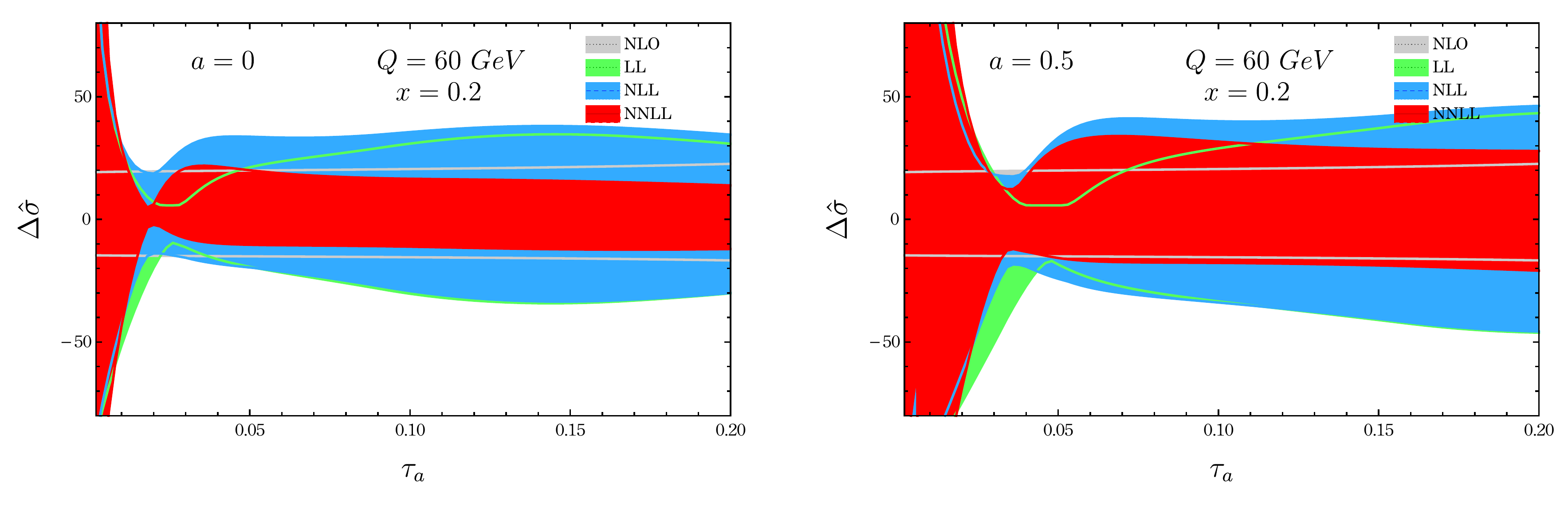} 
\end{center} 
\caption{Relative perturbative uncertainty in the differential cross-section for different $a=-1, -0.5, 0,$ and  $0.5$. }  \label{fig:Rdcsa} 
\end{figure}

In the numerical calculations we estimate the perturbative uncertainty generated by varying the scales $\mu_{H,S,J,B}$ given by the \textit{profile functions} presented in $e^+e^-$ angularity \cite{Bell:2018gce}. The profile functions are given in \appx{pro}. To estimate the uncertainties we specifically vary three parameters in the profile function \eq{scales} by a factor of 2 as $e_H=2^{\pm 1}, e_{S}= \pm 1/2$ and $e_{J,B}= \pm 1/2$. It is noticed that, the profile function in \cite{Bell:2018gce} is designed for $Z$-pole or higher scale and not so suitable for low $Q$ and large positive $a$ region-- show discontinuity in $\mu_s, \mu_{J,B}$ for $Q< 25\GeV$ and  $a>0.3$ region.  We discuss a possible modification in the profile function to access the the low $Q$ and positive $a$ in \appx{pro}. This modification provide access to the low $Q\sim 10\GeV$ and large angularity parameter $a=0.5$ region.   
For fixed-order uncertainty, $\mu$ is set to be $Q$ and varied up and down by a factor of two.

\begin{figure}[tbh]
\begin{center}
\includegraphics[scale=.5]{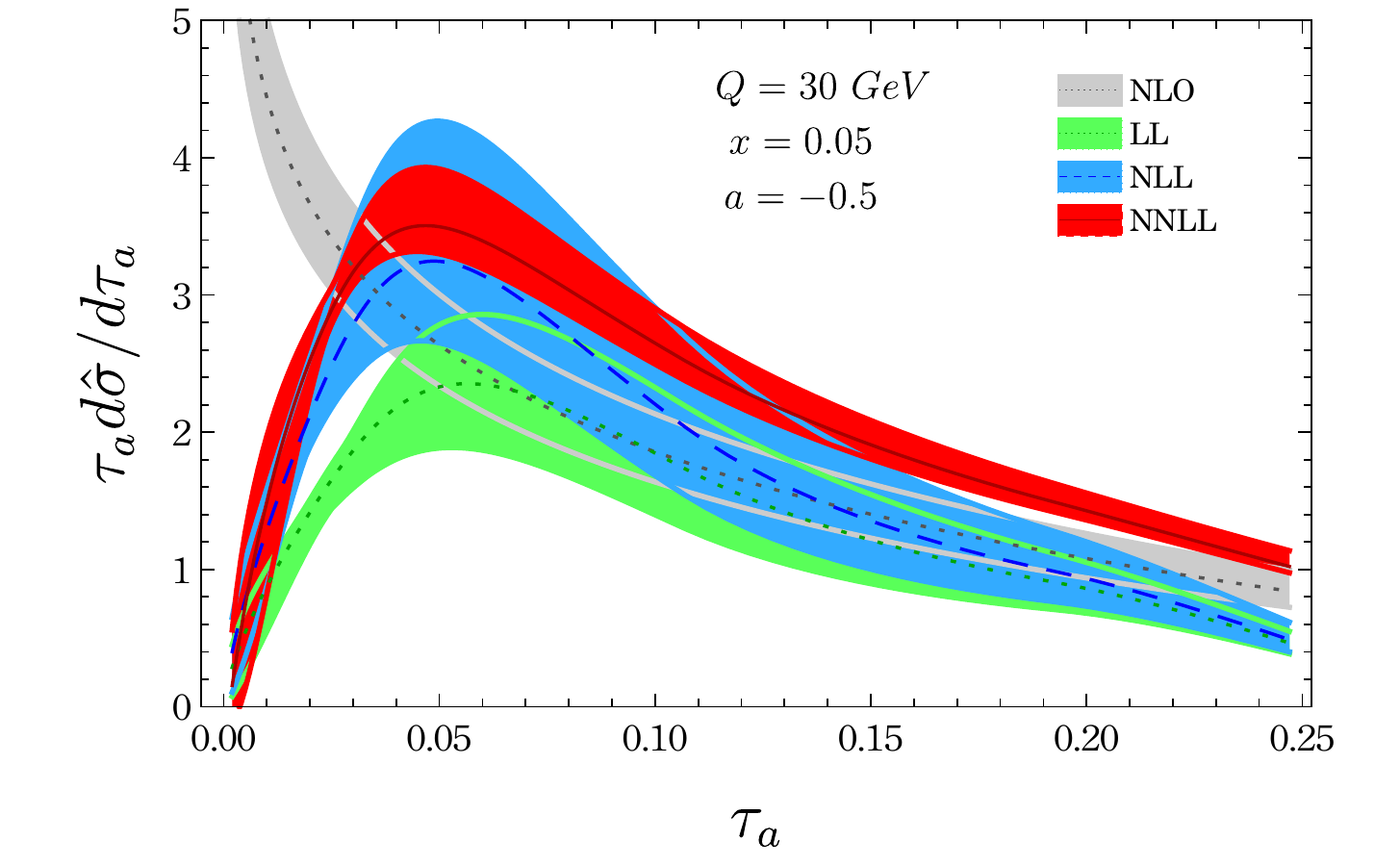} 
\includegraphics[scale=.5]{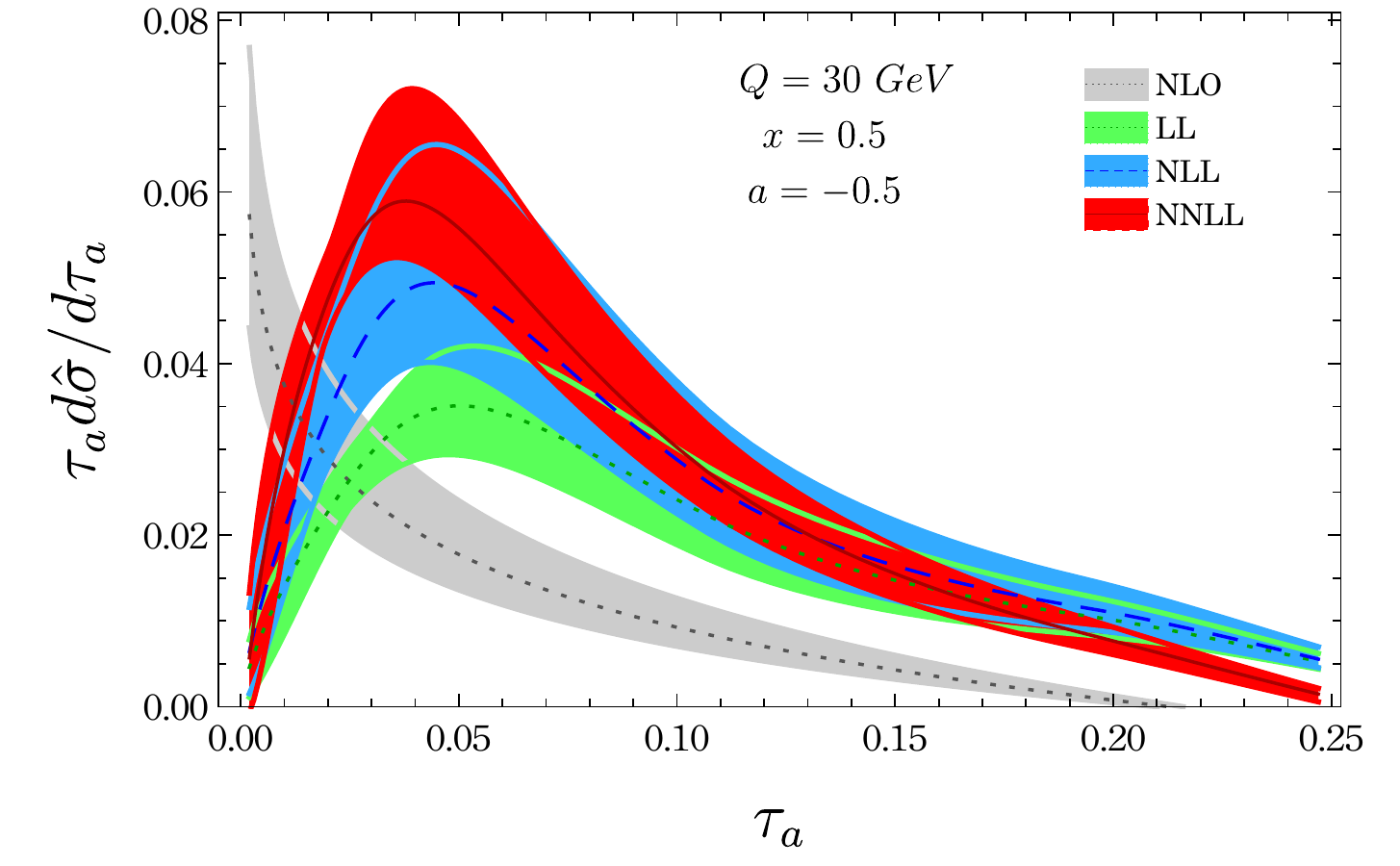}
\end{center} 
\caption{Angularity cross-section for different $x=0.05$, and $0.5$ at a fixed $a=-0.5 $ and $Q=30\GeV$. The band indicates perturbative uncertainty.} \label{fig:dcsx}
\end{figure}

\begin{figure}[tbh]
\begin{center}
\includegraphics[width=11.cm,height=4.2cm]{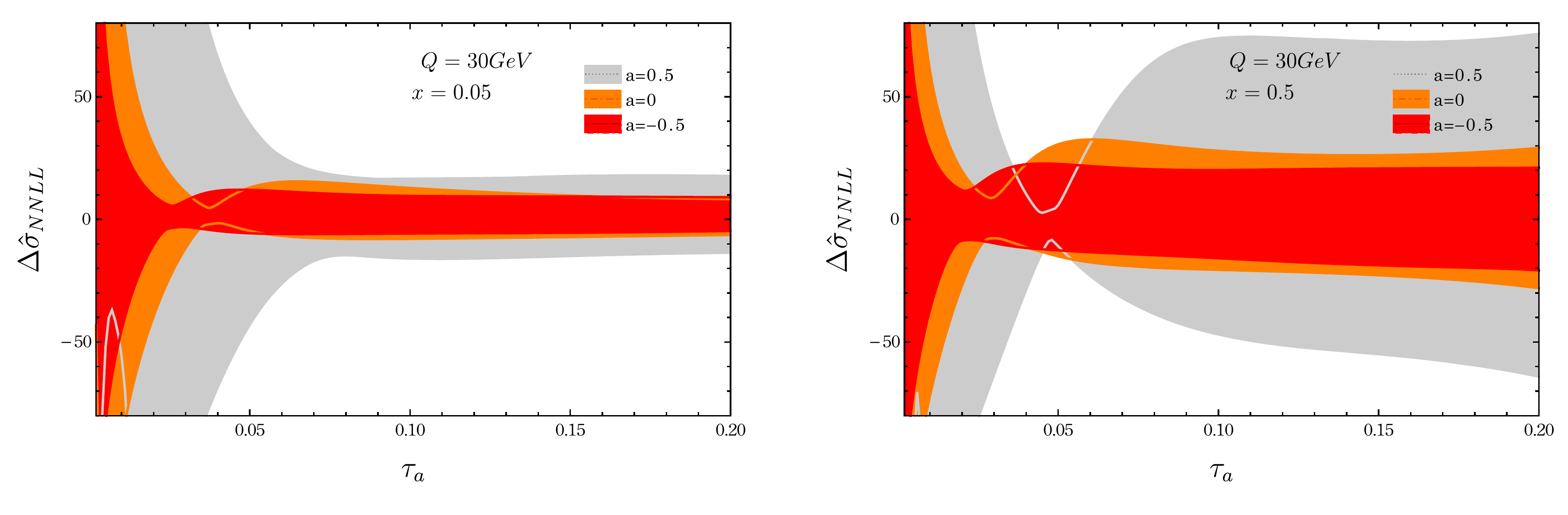} 
\end{center} 
\caption{Relative uncertainties at NNLL for different angularity $a$ dependence on different $x=0.05, 0.5$ and $a=-0.5, 0, 0.5$.} \label{fig:RdcsNNLL}
\end{figure}

\fig{dcsa} shows weighted differential cross-sections in \eq{hatsig} for different values of angularity parameter $a=-1,-0.5,0,0.5$ at a fixed $x=0.2$ and $Q=60$~GeV.  
Each sub-figure contains four plots: the NLO result with perturbative uncertainty is illustrated in gray and the resummed results at the LL, NLL and NNLL accuracy are shown in green, blue, and red curves with uncertainty bands respectively. The difference between the NLO and NNLL results shows the effect of the resummation. The whole cross-section can be characterized in three physical regions: the peak region ($\ta \sim 2 \Lambda_{QCD}/Q \ll 1$), the tail region ($ 2 \Lambda_{QCD}/Q \ll \ta  \ll 1$) and the far-tail region ($ \ta \sim 1$).
In the peak region, non-perturbative effect is dominated and our perturbative  approach is invalid. One can adopt a NP model \cite{Kang:2013nha} to take it into account. Since we do not include the NP effect and corresponding region $\ta\sim 0.01 $ in \fig{dcsa} should be modified significantly.
In the absence of NP effect, one sees that NLO result blows up due to singular terms $\ln(\ta)/\ta$, $1/\ta$, while the LL, NLL, NNLL results resum those terms and well-behave in the region. 
In the tail region, we find the resummation effect is still significant by comparing unresummed NLO and resummed NNLL.
The deviation of NLO from NNLL implies that conventional scale variations of NLO underestimates perturbative uncertainty of a pure fixed-order result in this region. 
We also find a reasonable overlapping between resummed results that imply perturbative convergence from LL to NNLL.  The NP effect in the region can be estimated as a power correction of a NP matrix element as shown in \cite{Lee:2006nr} and recently it is proposed that the power corrections can be determined in the framework of Eikonal  Dressed Gluon Exponentiation \cite{Agarwal:2020uxi}.
In the far tail region, 
the singular term are not dominated and resummation effect is not significant hence the nonsingular terms of $\as$, which are neglected in our calculation will give significant contributions. 
The region is not displayed in \fig{dcsa}.
The sub-figures in \fig{dcsa} indicate that the peak moves to right with the increasing value of $a$ and also the peak value increases. The results for $a=0$ represents the differential cross-section for DIS 1-jettiness with jet axis in \cite{Kang:2013nha}. 

For better understanding of uncertainty sensitivity on the parameters $a$, we present a relative percentage distributions in \fig{Rdcsa}.
The perturbative uncertainties increase with $a$  from 10\% to 30\% at NNLL accuracy in \fig{Rdcsa}.
The $a$ dependency is associated with the fact that the anomalous dimension in \eq{gamG} largely depends on $a$ through the constants $j_G$ and $\kappa_G$ multiplied by the cusp anomalous dimension. The large uncertainty at positive $a$ due to the factor $1/(1-a)$ in $\kappa_G$ and the singularity at $a=1$ implies that our result is invalid near that value as discussed in \sec{factorization}.  Once we manually change $a$ dependency in those constants while retaining the consistency in \eq{consis} still valid, we observed that the $a$ dependency in uncertainties accordingly changes. Therefore, the anomalous dimension sensitivity to $a$ leads to strong $a$ dependency in the perturbative uncertainty.

We also present the angularity cross-section for fixed $a=-0.5$ and different $x=0.05,  0.5$  in \fig{dcsx}, where the colors code are same as the colors code discussed above for \fig{dcsa}. The figures show that perturbative uncertainty increases with the increasing value of $x$. 
We also compare the uncertainty at NNLL for different $a=-0.5, 0, 0.5$ and low $x=0.05$ as well as for the high $x=0.5$ in \fig{RdcsNNLL}. The uncertainty (at NNLL) is minimum and less sensitive to $a$ for low $x=0.05$ unlike higher $x=0.5$.

\section{Conclusions}\label{sec:con}

In DIS, events with one final jet with initial state radiation dominate in final states at typical values of $x$. These events can be studied by using event shapes.
Angularity would be useful in DIS because it allows us to access a class of event shapes interpolating between thrust and broadening and extrapolating beyond the region by varying an angularity parameter $a$, and to perform event-shape analysis systematically in the $a$ space.

In this paper DIS angularity  $\ta$ is defined in terms of axes which take into account one-final jet and initial-state radiation along the proton-beam. The value of $\ta$ is small for such events and one can describe the region using soft and collinear degrees of freedom.
By using SCET we drive a factorization formula in $a<1$ region and it is composed of hard, beam, jet and soft functions. 
The jet and soft functions are the same as those appearing in $e^+e^-$ angularity and hard function appears in typical SCET factorization such as endpoint region or 1-jettiness in DIS. 

Our factorization contains an angularity beam function, which is  new result and we give the expression at $\mathcal{O}(\as)$.
The beam function has logarithmic terms which are known from jet function anomalous dimension
and newly presented constant terms, which are functions of $a$ as well as momentum fraction $x$ and are smoothly changing with $a$.
Our beam function at $a=0$ reproduces well-known thrust beam function. We compare the beam function with FJF which shares many terms in common. The difference between their perturbative matching coefficients is expressed in a single constant term proportional to the splitting function. We find that in positive $a$ region the difference increases relatively fast compared to negative $a$ region due to a contribution from FJF
while the contribution from beam function is steady in both regions.  We also show  $\ta$ distributions and perturbative uncertainties of beam function and find that they are sensitive to the value of $a$ while its generic behavior is similar to that of thrust beam function.

We also resum large logarithms of angularity by renormalization group evolution of the functions in our factorization 
and give our prediction for DIS angularity at next-to-next-to-leading log accuracy.
We adopted scale profile function used in $e^+e^-$ angularity and estimated perturbative uncertainties by varying
three scales in hard, jet, beam, and soft functions separately. We find the distributions are sensitive to the values of $a$ and perturbative uncertainties tends to be smaller in negative $a$ region but not further decreasing $a$ around $-1$ and below.

Studies of this paper include the factorization, beam function and resummed cross section for DIS angularity.
For application to future experiments such as EIC and/or EICC, there are several things to be considered in the future.
Our results contains singular terms and are correct in small $\ta$ region. Nonsingular terms obtained from fixed-order cross section will improve accuracy in large $\ta$ regions and also be useful to determine the scale profile function, which at this moment we borrow from that of $e^+e^-$ angularity.  Non-perturbative effect to angularity should be included in future studies.
For higher resummation, all ingredients except for two-loop constant of beam function are known to achieve so called NNLL$'$, which includes two-loop hard, beam, jet, and soft functions on top of NNLL accuracy and three-loop non-cusp anomalous dimension of soft or, jet function is necessary to achieve N$^3$LL accuracy.

\begin{acknowledgments}
The work of D.K. and T.M. is supported by the National Natural Science Foundation of China (NSFC) through Grant No. 11875112 and the China Postdoctoral Science Foundation through Grant No.KLH1512104. All authors contributed equally to this work.
\end{acknowledgments}
  
\appendix
\section*{APPENDIX}

\section{Intermediate steps of factorization}\label{app:interm}
In this section, we summarize intermediate steps 
where we perform 
summing over label indices,
applying label-momentum conservation,
and simplifying soft and collinear matrix elements 
to get the final form of factorization formula in \eq{fact}.

Now using matched QCD current onto SCET in \eq{matching} and decomposition of $\hat{\tau}_a$ in \eq{taua2}, the hadronic tensor of \eq{Wadef} can be written explicitly along with the matching currents and matching coefficients as 
\bea
&& W_{\mu\nu}(\ta) 
 =  \int d^4x \,e^{iq\cdot x} \sum_{\stackrel{\mbox{\scriptsize $n_1n_2$}}{n_1'n_2'}} 
\int d^3\tilde p_1 d^3\tilde p_2 d^3\tilde p_1' d^3 \tilde p_2' \, 
  e^{i(\tilde p_2 - \tilde p_1)\cdot x}  \int [d\ta ]  \bar C_{\mu}^{\beta\alpha}(\tilde p_1,\tilde p_2) \, 
  C_{\nu}^{\alpha'\beta'}(\tilde p_1',\tilde p_2') \,
 \nn \\
&&\quad \times \bra{P_{\nB}} \,\,
  \bar\chi_{n_2,\tilde p_2}^{\beta k}\overline T [Y_{n_2}^\dag Y_{n_1}]^{kj} \, 
  \chi_{n_1,\tilde p_1}^{\alpha j}(x) \,\, 
\hat\delta_J \hat\delta_B \hat\delta_S \,\,
 \bar\chi_{n_1',\tilde p_1'}^{\alpha' j'} T [Y_{n_1'}^\dag Y_{n_2'}]^{j'k'} \,
  \chi_{n_2',\tilde p_2'}^{\beta' k'}(0)  \ket{P_{\nB}}
\,,
\label{eq:Wlong}\eea
with  the short hand notations for integral 
$\int [ d\ta ]  =   \int d\ta^J d\ta^B d\ta^S \delta(\tau_a-\tau_a^J-\tau_a^B-\tau_a^{S})$
and measurement $\hat\delta_i=\delta(\ta^i - \hat{\tau}_a^{i})$.  The complex conjugate to the matching coefficient  $\bar C_{\mu}(\tilde p_1,\tilde p_2)=\gamma^0 C_{_\mu}^{\dagger}(\tilde p_1,\tilde p_2) \gamma^0$ and indeed it only depends on Lorentz scalar expressed as $C_{_\mu}^{\dagger}\bigl((\tilde p_1-\tilde p_2)^2\bigr)$.
The proton is considered as $n_B$-collinear state indicated and the state can decomposed into proton state in $n_B$ sector and vacuum states associated with independent sectors or, modes. In \eq{Wlong}, the collinear sectors (along $n_1$ and $n_2$) and the soft sectors  are decoupled and it can be factored out in terms of collinear and soft matrix elements of associated vacuum states.  By taking into account that the fields within each collinear matrix element are zero if they are not aligned with the collinear sector, we sum over all $n_{1,2}$ and $n^\prime_{1,2}$ and come up with $n_J$ or $n_B$. Now absorbing the integrals over $\tilde{p}_{1,2}$ into the unlabeled fields $\chi_{n_{1,2}}$ and decomposing the proton matrix element into the soft and collinear matrix elements one can write the hadronic tensor as
\bea
W_{\mu\nu}(\ta) &=&  \int d^4 x\int d^3\tilde p_1 d^3\tilde p_2 
  e^{i(q + \tilde p_2 - \tilde p_1)\cdot x}  \int [d\ta ] 
  \bar C_{ \mu}^{\beta\alpha}((\tilde p_1-\tilde p_2)^2) \,
  C_{ \nu}^{\alpha'\beta'}((\tilde p_1-\tilde p_2)^2) 
 \nn\\
& \times& \bra{0}[Y_{\nB}^\dag Y_{\nJ}]^{kj}(x) \,\,
 \hat\delta_S\,\,   
  [Y_{\nJ}^\dag Y_{\nB}]^{j'k'}(0)\ket{0}
 \nn\\
& \times&  \biggl\{ \! \bra{P_{\nB}} \bar\chi_{\nB,\tilde p_2}^{\beta k}(x) 
 \,\,\hat\delta_B\,\,
 \chi_{\nB}^{\beta' k'}(0) \ket{P_{\nB}} \bra{0}\chi_{\nJ,\tilde p_1}^{\alpha j}(x)  \,\,\hat\delta_J\,\, 
  \bar\chi_{\nJ}^{\alpha' j'}(0)\ket{0} 
+(\chi \leftrightarrow \bar\chi) \biggr\}
 \,.
 \label{eq:Wfactored}\eea
Since the Wilson lines are space-like separated and the time ordering is same as the path ordering \cite{Bauer:2003di,Fleming:2007qr}, we use $T [Y_{n_J}^\dag Y_{n_B}]=  [Y_{n_J}^\dag Y_{n_B}]$ and $\overline{T} [Y_{n_J}^\dag Y_{n_B}]=  [Y_{n_J}^\dag Y_{n_B}]$ in the soft matrix element. Two terms in the last line come from the fact that we have two ways to chose a pair of collinear fields in the proton matrix element and they turn into quark and anti-quark beam and jet functions. 
 
Now let us perform the $x$ integration. During the integration we can ignore $x$ dependence in the collinear field $\chi_{n,\tilde{p}}(x)$ and soft Wilson line  $Y_n(x)$ because these fields describe fluctuations of low energy/momentum modes and in exponent terms like $e^{ik_s\cdot x}$ or, $e^{i n\cdot k\, \bn\mcdot x/2}$ of these fields, the momenta are of residual scale  suppressed compared to label or hard momenta in the exponent in first line of \eq{Wfactored}.

By expressing dot product in $x\cdot q$, $x\cdot \tilde{p}_1$, $x\cdot \tilde{p}_2$ in terms of $n_B,\bar{n}_B$ or,  $n_J,\bar{n}_J$ coordinates in \eqs{nBJ}{bnBJ}
\be
(q+\tilde p_2 - \tilde p_1)\cdot x
 =
 (\bnB\mcdot q + \omega_2)\frac{\nB\mcdot x}{2}
 +(\bnJ\mcdot q - \omega_1)\frac{\nJ\mcdot x}{2}
 +(q_\perp +\tilde p_2^\perp - \tilde p_1^\perp)\cdot x_\perp
\,.
\label{eq:pcons}\ee
Note that $\omega_{1,2}$ are large component of the label momentum: $\tilde{p}_{i}=\frac{\omega_i n_i}{2} + \tilde{p}^\perp_i $.
Rewriting the differential 
$d^4 x=\tfrac{1}{2}d(\nB\mcdot x)\, d(\bnB\mcdot x)\, d^2 x_\perp$
we obtain
\bea
\int d^4 x \, e^{i(q+\tilde p_2 - \tilde p_1)\cdot x} 
&=&\delta^{(4)}(q+\tilde p_2 - \tilde p_1)
\nn\\
&=&
\frac{4}{\nJ\mcdot \nB}(2\pi)^4
\delta(\bnB\mcdot q + \omega_2) \,
 \delta(\bnJ\mcdot q -\omega_1) \,
 \delta^{(2)}(q_\perp +\tilde p_2^\perp -\tilde p_1^\perp)
 \,.\label{eq:delta1}
\eea

Then, we have
\bea
\tilde p_1 &=&\bnJ\mcdot q \frac{\nJ}{2}+q_\perp+\tilde p_\perp
=\bnJ\mcdot q \frac{\nJ}{2}
\nn\\
\tilde p_2 &=& -\bnB\mcdot q \frac{\nB}{2}+\tilde p_\perp
\,,
\label{eq:p1p2}
\eea
where we dropped subscript in $\tilde p_{2\perp}=\tilde p_{\perp}$ and in the first line, we set the transverse momentum to zero $q_\perp+\tilde p_\perp=0$ according to definition of the jet axis $\nJ$ which we adjust such that it is aligned with jet momentum and no transverse momentum is left in $\tilde p_1$.
The hadronic tensor reads as
\bea 
W_{\mu \nu}(\ta) 
&=& \frac{4(2\pi)^4}{n_B.n_J}
    \int d^2\tilde{p}_\perp  
    \int [d\ta]
\nn \\
&&\times
\bar{C}^{\beta\alpha}_{\mu}(q^2) \, C^{\alpha' \beta'}_{\nu}(q^2)
\nn\\
&& \times \langle 0 | [Y^\dagger_{n_B}Y_{n_J}]^{kj}(0) \,\hat\delta_S\,
[Y^\dagger_{n_J} Y_{n_B}]^{j'k'}(0)| 0 \rangle 
\nn \\
&& \times \bigg\lbrace \langle P_{n_B}| \bar{\chi}^{\beta k}_{n_B}(0)
\,\hat\delta_B\, 
\Bigr[\delta(\bar{n}_B\mcdot q+\bar{n}_B\mcdot \mathcal{P}) \delta^{(2)}(\tilde{p}_\perp - \mathcal{P}_\perp) \chi^{\beta' k'}_{n_B}(0)  \Bigr] | P _ {n_B} \rangle 
\nn \\
&& \times \langle 0| \chi^{\alpha j}_{n_J}(0)
\,\hat\delta_J\, 
\Bigl[\delta(\bar{n}_J\mcdot q+\bar{n}_J\mcdot \mathcal{P}) \delta^{(2)}(\mathcal{P}_\perp) \bar{\chi}^{\alpha' j'}_{n_J}(0)  \Bigr] | 0\rangle \nn\\
&&+(\chi\leftrightarrow\bar \chi) \bigg\rbrace
\nn \\
&=& \frac{4(2\pi)^4}{n_B\cdot n_J} \bar{C}^{\beta\alpha}_{\mu}(q^2) \, C^{\alpha'\beta'}_{\nu}(q^2)
    \int [d\ta]
\nn\\
&& \times \langle 0 | [Y^\dagger_{n_B}Y_{n_J}]^{kj}(0) \,\hat\delta_S\,
[Y^\dagger_{n_J} Y_{n_B}]^{j'k'}(0)| 0 \rangle 
\nn \\
&& \times \bigg\lbrace \langle P_{n_B}| \bar{\chi}^{\beta k}_{n_B}(0)
\,\hat\delta_B\, 
\Bigr[\delta(\bar{n}_B\mcdot q+\bar{n}_B\mcdot \mathcal{P}) \chi^{\beta' k'}_{n_B}(0)  \Bigr] | P _ {n_B} \rangle 
\nn \\
&& \times
\langle 0| \chi^{\alpha j}_{n_J}(0)
\,\hat\delta_J\,
\Bigl[\delta(\bar{n}_J\mcdot q+\bar{n}_J\mcdot \mathcal{P}) \delta^{(2)}(\mathcal{P}_\perp) \bar{\chi}^{\alpha' j'}_{n_J}(0)  \Bigr] | 0\rangle \nn\\
&&+(\chi\leftrightarrow\bar \chi) \bigg\rbrace
\,,\label{eq:W_facto}
\eea
where in the second equality, the matching coefficients $C$ depending on $q^2$ is moved out of the integral and transverse momentum delta function in the proton matrix element is integrated. Now one soft and two collinear matrix elements are convolved by $\ta$ integrals  and corresponding measurement functions $\hat \delta_{S,J,B}$.
Matrix elements in the above expression can be related to the known hard, soft, jet, and beam functions and by using these functions we further simplify the expression.

Next we express the matrix elements of \eq{W_facto} in terms of the hard, jet, soft and beam function defined in the SCET formalism.
The collinear vacuum matrix elements in \eq{W_facto} can be expressed in terms of the angularity jet function \cite{Hornig:2009vb}. The jet function integrating over $\ell^+$ and over $x$ in Eq.~(2.17a)\cite{Hornig:2009vb} is given by
\bea
J_q(e_n,w)\bigg(\frac{\slashed{n}}{2}\bigg)_{\alpha\beta}= \frac{2}{N_c} (2\pi)^3 \text{Tr} \langle 0| \chi^{\alpha}_{n}(0)\delta(e_n-\hat{e}_n) \delta(w+\bar{n}\cdot \mathcal{P}) \delta^{(2)}(\mathcal{P}_\perp) \bar{\chi}^{\beta}_{n}(0) | 0\rangle 
\,,  \label{eq:J_def}
\eea
where we specify large momentum component $w$ in the argument.
The $e^+e^-$ angularity defined in \cite{Hornig:2009vb} differs from DIS angularity $\hat{\tau}^J_a$ in normalization and for a particle of momentum $p$, they are related
\bea 
e_n = \frac{1}{w} (n\cdot p)^{1-\frac a2}(\bn \cdot p)^{\frac a2}=A^{-1} \tau_a
\label{eq:en}
\,,
\eea
where we find $w=\bnJ \cdot q$ by comparing \eq{J_def} and \eq{W_facto}. 
\bea
A=\frac{w w_J^{1-a} (2\qB\mcdot \qJ)^{\frac a2}}{Q^2}
=\left(\frac{w_J}{\sqrt{2 q_B\cdot q_J}}\right)^{2-a}
=\left(\frac{w_J}{Q}\right)^{2-a}+O(\lambda^2)
\,,\label{eq:A}
\eea
where we use approximations $w = Q^2 w_J/(2 \qB\mcdot\qJ)$ and $2\qB\cdot \qJ\approx Q^2$, which are correct up to power corrections of order $\lambda^2$.
Then, the jet function that measures $\ta$ can be rewritten in terms of \eq{J_def} after rescaling of the measurement delta function by $A$.
\bea
\delta(\ta^J-\hat{\ta}^J)=A^{-1}\delta(A^{-1}\ta^J-\hat{e}_n)
\,.
\label{eq:dtaJ}\eea
Now the collinear matrix element for vacuum state in \eq{W_facto} can be expressed as
\bea 
\langle 0| \chi^{\alpha j}_{n_J}(0)\delta(\tau_a^{J}-\hat{\tau}_a^{J})&& \delta(\bnJ\cdot q+\bar{n}_J\cdot\mathcal{P}) \delta^{(2)}({P}_\perp) \bar{\chi}^{\alpha' j'}_{n_J}(0) | 0\rangle 
\nn\\
&&=\frac{\delta^{j j'} }{(2\pi)^3}\bigg(\frac{{\slashed{n}_J}}{4}\bigg)_{\alpha\alpha'} 
\,\, A^{-1} J_q(A^{-1}\tau_a^J, \bnJ\mcdot q)
\nn\\
&&=\frac{\delta^{j j'} }{(2\pi)^3}\bigg(\frac{{\slashed{n}_J}}{4}\bigg)_{\alpha\alpha'} 
J_q(\tau_a^J, Q)
\,
\label{eq:J_Mq} \eea
in the last step we used the one-loop expression in \cite{Hornig:2009vb} and identified that a term of the form $\mu/(\bnJ\mcdot q)$ becomes $A^{1/(2-a)}\mu/(\bnJ\mcdot q)=\mu\sqrt{2 q_B\mcdot q_J}/Q^2=\mu/Q+\mathcal{O}(\lambda^2)$.
From now on, we drop flavor index $q$ since we have just quark-jet function, which is the same for any flavor $q$. Similarly for the soft function we also omit the index.

For the proton matrix element we can define a angularity beam function similar to jet function definition:
\be 
\mathcal{B}_q(e_n, x;w) =                                                 \theta(w) \langle P_{n}| \bar{\chi}_n(0)\frac{\bar{\slashed{n}}}{2}\delta(e_n-\hat{e}_n) \Bigr[\delta(w-\bn\cdot \mathcal{P}) \chi_{n}(0)  \Bigr] | P_{n} \rangle 
\,.
\label{eq:B_def}\ee
We have the similar relation and factor as in \eqs{en}{A} with $w=-\bnB\cdot q=-q_J\cdot q  w_B/(q_B\cdot q_J)$ and $A=(w_B/Q)^{2-a}+\mathcal{O}(\lambda)$ where we use $q_J\cdot q =(xP+q)\cdot q [1+ \mathcal{O}(\lambda)]=-Q^2/2$.
The same holds for measurement function in \eq{dtaJ} with replacement of $\ta^J$ by $\ta^B$.

The collinear proton matrix elements in \eq{W_facto} can be written in terms of the angularity beam function of \eq{B_def} as 
\bea 
&&\langle P_{n_B}| \bar{\chi}^{\beta k}_{n_B}(0)
\delta(\ta^{B}-\hat{\tau}_a^{B}) \Bigr[\delta(\bnB\mcdot q+\bnB\mcdot\cP) \chi^{\beta' k'}_{n_B}(0)  \Bigr] | P_{n_B}\rangle 
\nn \\
&&=\frac{\delta^{k k'}}{N_c}\bigg(\frac{\slashed{n}_B}{4}\bigg)_{\beta'\beta}  A^{-1} \cB_q(A^{-1}\tau_a^B, x, -\bnB\mcdot q)
\nn\\
&&=\frac{\delta^{k k'}}{N_c}\bigg(\frac{\slashed{n}_B}{4}\bigg)_{\beta'\beta}  \cB_q(\ta^B, x, Q)
\,,
\label{eq:B_Mq} \eea
Where in the last step, we used the one-loop expression in \eq{cB1q} and the scaling factor $A$ is absorbed in the way that replaces $-\bnB\cdot q$ by $Q$ as in \eq{J_Mq}.

The angularity soft function was obtained in \cite{Hornig:2009vb}
\be
S(e_s, Q) = \frac{1}{N_c} 
Tr \langle 0|  \bar{Y}^\dagger_{\bn} Y^\dagger_n (0)\, 
\delta(e_s-\hat{e}_s  )  Y_n \bar{Y}_{\bn} (0)| 0 \rangle|
\,.
\label{eq:Ses}\ee
It is expressed with Wilson lines along $n$ and along $\bn$ directions with a normalization $n\mcdot \bn =2$ while  $\nJ$ and $\nB$  of Wilson lines in \eq{W_facto} is different $n_J\mcdot n_B \neq 2$. We rescale them and rewrite soft matrix elements  in \eq{W_facto} in terms of
\be 
n' =\nB\frac{w_B}{\sqrt{2 q_B\mcdot q_J}} 
\, ,\quad \bn'=\nJ\frac{w_J}{\sqrt{2 q_B\mcdot q_J}}
\quad
\text{and}\, ,\quad
n'\cdot \bn'=2
\,.
\label{eq:np}\ee
The Wilson line $Y_n$ is invariant under re-scaling of $n$ by a constant factor $\alpha$:  $Y_{n}=Y_{\alpha n}$ then, we have $Y_{n_B}=Y_{n'}$ and $Y_{n_J}=Y_{\bn'}$. 

When a particle enters the hemisphere $n\mcdot p < \bn \mcdot p$  the observable $e_s$ is given by
\be
e_s =\frac{1}{Q}(n\mcdot p)^{1-\frac a2}(\bn \mcdot p)^\frac{a}{2}
\,,
\label{eq:es}
\ee
While our angularity for $q_B\mcdot p <q_J\mcdot p$ equivalently, $n'\mcdot p <\bn' \mcdot p$ is given
\be 
\ta^S=\frac{2}{Q^2}(q_B\mcdot p)^{1-\frac a2}(q_J\mcdot p)^\frac{a}{2}
=\frac{1}{Q_R} (n'\mcdot p)^{(1-\frac a2)}(\bn' \mcdot p)^\frac{a}{2}
=A e_s
\,,
\label{eq:taS}\ee
where $Q_R= Q^2/\sqrt{2q_B\mcdot q_J}$ and  $A=Q/Q_R$ is the scale factor for the soft function.
This is still valid for multi-particle final state. 

The soft matrix element from \eq{W_facto} can be expressed in terms of angularity soft function as 
\bea
&&
\langle 0 | \Bigr[Y^\dagger_{n_B}Y_{n_J}\Bigr]^{kj}(0) 
\delta(\tau_a^{S}-\hat{\tau}_a^{S})
\Bigr[Y^\dagger_{n_B}Y_{n_J}\Bigr]^{j'k'}(0)| 0 \rangle 
\nn \\
&&
=\langle 0 | \Bigr[Y^\dagger_{n'}Y_{\bn}\Bigr]^{kj}(0) 
A^{-1}\delta(A^{-1}\tau_a^{S}-e_s)
\Bigr[Y^\dagger_{n'}Y_{\bn'}\Bigr]^{j'k'}(0)| 0 \rangle 
\nn\\
&&
=\frac{\delta^{jj'}\delta^{kk'}}{N_c}  A^{-1} S(A^{-1}\ta^S,Q)
\nn\\
&&
=\frac{\delta^{jj'}\delta^{kk'}}{N_c}  S(\ta^S, Q_R)
\,, \label{eq:S_mat}\eea
where we again can use $Q_R=Q[1+\mathcal{O}(\lambda)]$.
In the end, we come up with simple rescaling results of jet, beam, and soft functions from $e^+e^-$ angularity to DIS angularity. \eqss{J_Mq}{B_Mq}{S_mat} show that we simply set second argument of each function to be $Q$ up to power corrections of $\mathcal{O}(\lambda)$. From now on, we make it implicit in argument of the function if unnecessary and pretend $Q$ is the large momentum component of all functions as is CM frame in $e^+e^-$ annihilation. On the other hand, we make the scale $\mu$ dependence expressed in the argument of functions in \eqs{Wfact}{fact}. 
 
We are ready to insert \eqss{J_Mq}{B_Mq}{S_mat} to simplify \eq{W_facto}.
In doing so the matching coefficient is traced 
\bea 
H_{\mu\nu}(q^2)
&=&\text{Tr} \left( \frac{\slashed{n}_B}{4} \bar{C}_\mu (q^2) \frac{\slashed{n}_J}{4} C_\nu (q^2) \right)
\nn\\
&=&\text{Tr} \left( \frac{\slashed{n}_B}{4} \gamma^{\perp}_\mu \frac{\slashed{n}_J}{4} \gamma^{\perp}_\nu \right)
|C(q^2)|^2
\nn\\
&=&
\frac{\nB\mcdot\nJ}{4} |C(q^2)|^2 \, 
\left( -g_{\mu\nu}+\frac{\qB^\mu\qJ^\nu +\qB^\nu\qJ^\mu}{\qB\cdot\qJ}
\right)
\,,
\label{eq:TrCC}\eea 
where we rewrote the matching coefficient $C_\mu (q^2)=\gamma^\perp_\mu C(q^2)$, where $C(q^2)$ is a scalar function and $\bar{C}_\mu=\gamma^0 C^\dagger_\mu(q^2)\gamma^0=\gamma^\perp_\mu C^\dagger (q^2) $.
We finally obtain the expression in \eq{Wfact}.

The hadronic tensor in \eq{Wfact} is contracted with the lepton tensor in \eq{lepdef}
\be
 L^{\mu \nu }(x,Q^2) \frac{8 \pi}{n_J.n_B} H_{\mu \nu}(q^2) 
=\frac{d\sigma_0}{dx dQ^2}  H(Q^2,\mu)
\,,\label{eq:LH}\ee
where the hard function is defined as
\bea 
H_f(Q^2) = Q_f^2 \,|C(q^2)|^2
\label{eq:H}\eea 
and the born level cross section is given by
\bea
\frac{d\sigma_0}{dx dQ^2}
&=&\frac{4\pi \alpha_\text{em}^2 }{2x^2 s^2 Q^2}
\frac{\qB\mcdot k\qJ\mcdot k' +\qB\mcdot k' \qJ\mcdot k}{\qB\cdot\qJ}
\nn\\
  &=&\frac{ 2\pi\alpha^2_{\text{em}} }{Q^{4}}[ (1-y)^2 + 1] \left( 1+ \mathcal{O}(\lambda)\right)
\,,
\label{eq:born} \eea 
in the last step we use approximation $\qJ=x P+q + \mathcal{O}(\lambda Q)$.

\section{Beam function computation} \label{app:oneloop}
Here we compute the constant terms $\wtc_1^{qq}$ and   $\wtc_1^{qg}$ in the one-loop correction  $\tcI^{(1)}_{qj}$ given in \eq{tI1}. 
The bare beam function for beam thrust is presented in several works  \cite{Stewart:2010qs,Jain:2011iu,Gaunt:2014xga,Gaunt:2014xxa,Ritzmann:2014mka}.
At one-loop, we just have a single parton emission from initial parton and by using relation between thrust and angularity for a single particle we can obtain the angularity beam function from known thrust beam function in \cite{Ritzmann:2014mka}. 
 
In Breit frame, the initial parton with momentum $P^-=Q/z$ splits into struck quark with the large  momentum component $z P^-=Q$ and the emitted gluon with $(1-z)P^-=\tfrac{1-z}{z}Q$. The angularity is written in terms of the gluon momentum $k$ as
\be
\ta^B=\frac{ (k^+)^{1-a/2}(k^-)^{a/2}}{Q}=\left(\frac{1-z}{z}\right)^{a/2} \left(\frac{k^+}{Q}\right)^{1-a/2}
\,.
\label{eq:tabeam}
\ee
Note that, we showed in \appx{interm} that this is equivalent to the definition given in \eq{tau_df} up to corrections suppressed by powers of $\lambda$. For a single particle, the angularity $\ta^B$ is related to the beam thrust $t$ as
\be
t=Q k^+= Q^2 \ta^{\frac{2}{2-a}}\left(\frac{z}{1-z}\right)^\frac{a}{2-a}
\,.
\label{eq:ta-beam} \ee
Now we rewriting the one-loop expression given in \cite{Ritzmann:2014mka} in terms of angularity as
\be
\cB_q^\text{bare}=\frac{\as C_F}{2\pi}
 \frac{2}{2-a} \left(1-\e^2\frac{\pi^2}{12}\right) 
\left(\frac{\mu^2}{Q^2}\right)^\e \left( \frac{1}{\ta}\right)^{1+\tfrac{2\e}{2-a}} h_q(z,\e)
\,,
\label{eq:cB1q}\ee
where the function $h_q$ is given by
\be
h_q(z,\e)= z^{\e\fraca} \left[ \frac{1+z^2}{(1-z)^{1+\e\fraca}} -\e(1-z)^{1-\e\fraca} \right] 
\,.
\label{eq:fq}\ee

In the Laplace space, quark beam function  $\tcB_{q/P}(\nu,z,\mu)$ is factorized into the Laplace-space coefficient $\tcI_{qj}(\nu, z/\xi,\mu)$ and proton PDF $f_{j/P}(z)$, where $\nu$ is conjugate variable to $\ta$ as in \eq{tBIf}. The perturbative beam function is computed for quark or gluon state instead of proton. At one-loop, the function $\tcB_{q/k}$ for a parton $k=q,g$ can be written as
\be
\tcB_{q/k}(\nu,z,\mu)=\sum_{j=q,g}\tcI_{qj} \otimes f_{j/k}
\,.
\label{eq:tBqk}\ee
By comparing \eqs{tBqk}{tBIf} one finds proton PDF $f_j$ is replaced by the parton PDF $f_{j/k}$ with a flavor $k$ while the matching coefficient $\tcI_{qj}$ remains the same.
The parton PDFs to order $\as$ are 
\be
f^{(0)}_{j/k}= \mathbbm{1}_{jk}\,,
\qquad
\qquad
f^{(1)}_{j/k}=-\frac{1}{\e}\frac{\as C_{jk}}{2\pi} P_{jk} (z)
\,,
\label{eq:f0f1}\ee
where $\mathbbm{1}_{jk}$ is defined below \eq{In}, the color factors are $C_{qq,\bar q \bar q}=C_F$ and $C_{qg,\bar q g}=T_F$, and the splitting functions $P_{ik}$ are given in \eq{PqqPqg}.
Therefore, one-loop contribution from parton can be written as
\be
\tcB_{q/k}^{(1)} 
= \tcI^{(1)}_{qk}-\frac{1}{\e}\frac{\as C_{qk}}{2\pi}  P_{qk}(z)
\,.
\label{eq:wtB1}\ee
The above expression is compared to renormalized beam function and the coefficients $\tcI_{qq,qg}$ are determined.

The term $1/\ta^{1+\e}$ in the bare function \eq{cB1q} turns into  $\Gamma(-\e) \, \nu^\e$  by the Laplace transformation defined in \eq{LP}. 
The bare function in the Laplace space is given by
\be \label{eq:tcB1q}
\wt{\cB}^\text{bare}_q=\frac{\as C_F}{2\pi}
 \frac{2}{2-a} \left(1-\e^2\frac{\pi^2}{12}\right)
 \Gamma\left[-\frac{2\e}{2-a}\right]
 \left(\frac{Q\nu^{-\frac{1}{2-a} }}{\mu} \right)^{-2\e} h_q(z,\e)
\,.\ee
After expanding \eq{tcB1q} in powers of $\e$, 
we identify IR divergence that is matched to one-loop quark PDF in \eq{f0f1}.  
The beam function have the same IR divergence with PDF \cite{Stewart:2010qs} and
leftover divergences are UV and they are absorbed into a renormalization factor $\wt{Z}_{qq}$. 
The finite term is the matching coefficient $\tcI_{qq}$. In the case of jet function, all IR divergences are cancelled as shown in \cite{Hornig:2009vb}.
Up to finite terms, we have
\bea
\wt{\cB}^\text{bare}_q &=& \wt{Z}^{(1)}_{qq}(\e)-\frac{1}{\e}\frac{\as C_F}{2\pi} P_{qq}(z) +\tcI_{qq}^{(1)}(\nu,z)
\,,
\nn\\
 \wt{Z}^{(1)}_{qq}(\e)&=&\frac{\as C_F}{2\pi}\left\{\left[\frac{1}{\e^2}-\frac{2}{\e}L_B\right]  \frac{2-a}{1-a}+\frac{3}{2}\frac{1}{\e} \right\} \delta(1-z)
\,.
\label{eq:wtBwtZ}
\eea

The finite term at one loop is the matching coefficient with the form
\bea
\tcI_{qq}^{(1)}(\nu,z) 
&=& \frac{\as}{4\pi}\left\{ 4 C_F\left[\frac{2-a}{1-a} \delta(1-z) L_B^2 +  \left(P_{qq}(z)-\frac{3}{2}\delta(1-z)\right) L_B \right]+ \wt{c}_1^{qq}(z,a)\right\}
\,,
\label{eq:I1qq} 
\eea
where the constant term $ \wt{c}_1^{qq}$ is given by \eq{wtc1qq}.
The constant $\wt{c}_1^{qq}$ is newly computed for the first time while the coefficients of $L_B^2$ and $L_B$ are the anomalous dimensions known from jet function which can be found easily comparing to \eq{Gfo}.

The beam functions for gluon initial state in momentum space  \cite{Ritzmann:2014mka} and Laplace space are given by 
\bea
\cB_g^\text{bare}&=&\frac{\as T_F}{2\pi}
 \frac{2}{2-a} (1+\e)  \left(\frac{\mu^2}{Q^2}\right)^\e \left( \frac{1}{\ta}\right)^{1+\tfrac{2\e}{2-a}} h_g(z,\e),
\nn\\
\wt{\cB}_g^\text{bare}&=& \frac{\as T_F}{2\pi} 
 \frac{2}{2-a} (1+\e) \Gamma\left[-\frac{2\e}{2-a}\right]
 \left(\frac{Q\nu^{-\frac{1}{2-a} }}{\mu} \right)^{-2\e} h_g(z,\e),
\nn\\
h_g(z,\e) &=& z^{\e\fraca} \left[ \frac{z^2-\e}{ (1-z)^{\e\fraca}}+(1-z)^{2-\e\fraca}\right] 
\,.
\label{eq:BBhg}\eea
The bare function matches to bare gluon PDF and the matching coefficient as
\be
\wt{\cB}_g^\text{bare}=-\frac{1}{\e}\frac{\as T_F}{2\pi}P_{qg}(z)
+\tcI_{qg}^{(1)}(\nu, z),
\label{eq:Bg_match}\ee
where the matching coefficient is given by
\bea
 \tcI_{qg}^{(1)}(z,\nu) = \frac{\as }{4\pi} 
\left[ 4T_F P_{qg}(z) L_B+ \wtc_1^{qg}(z)\right]
\label{eq:I1qg}\eea
and the constant $\wtc_1^{qg}$ is given by \eq{wtc1qg}.

For completeness we give the one-loop result in momentum space. By using the identity $1/\ta^{(1+\e)} \leftrightarrow \Gamma(-\e)/\nu^\e$ between momentum and Laplace space and expanding in powers of $\e$ we have
\bea
\delta(\tau) &\leftrightarrow & 1
\nn\\
\cL_0(\tau) &\leftrightarrow & -\ln\left(\nu e^{\gamma_E}\right)
\nn\\
\cL_1(\tau) &\leftrightarrow & \frac{1}{2}\ln^2\left(\nu e^{\gamma_E}\right)+\frac{\pi^2}{12}  
\label{eq:cLnLn}\eea

We obtain
\bea
\cI_{qq}^{(1)}(\ta,z,\mu) &=& \frac{\as C_F}{2\pi}
 \left\{ \frac{4\delta(1-z) }{(2-a)(1-a)} \cL_1(\ta)+ \right.
 \nn\\
  && \qquad \qquad  \left.
 \left[\frac{2}{2-a}\left(P_{qq}(z)-\frac{3}{2}\delta(1-z)\right)  - \frac{2}{1-a} \delta(1-z) L_\mu  \right]\cL_0(\ta)\right. 
 \nn\\ 
  && \qquad \qquad  \left.
 	+\left[- (1+z^2) \cL_0(1-z) L_\mu +\frac{2-a}{2(1-a)}\delta(1-z) L_\mu^2\right]\cL_{-1} (\ta)\right\}		
\nn\\
  && \qquad \qquad  
	+  \frac{\as}{4\pi}c^1_{qq}(z,a) \cL_{-1} (\ta) \label{eq:Iqq1tau}\\
c^1_{qq}(z,a) &=& \wt{c}^1_{qq}(z,a) -C_F\frac{2\pi^2}{3}\frac{\delta(1-z)}{(1-a)(2-a)}
\,,
\label{eq:I1c1qq}\eea

where $L_\mu= \ln (\mu^2/Q^2)$.
\bea
\cI_{qg}^{(1)}(\ta,z,\mu) &=& \frac{\as }{4\pi} 
\left\{ 2T_F\left(\frac{2}{2-a} P_{qg}(z)\cL_0(\ta) -P_{qg}(z) L_\mu)\right)+ c^1_{qg}(z,a)\cL_{-1}(\ta) \right\} \label{eq:Iqg1tau}\\
c^1_{qg}(z,a) &=&\wt{c}^1_{qg}(z,a)
\,.
\label{eq:I1c1qg}\eea

\section{Profile functions in high and low $Q$ regions} \label{app:pro}
For completeness, we rewrite the adopted profile functions presented for $e^+e^-$ angularity in  \cite{Bell:2018gce}, which is aimed for high $Q$ region around the $Z$ pole and above, and needs a proper modification to apply for lower $Q$ region. We discuss the modification to access the low $Q$ and large positive $a$ region in detail.  

The factorized cross-section in \eq{sigta} has four scales-- hard scale $\mu_H$, soft scale $\mu_S$, jet scale $\mu_J$ and beam scale $\mu_B$, each of which is related to corresponding hard, soft, jet, and beam functions. By choosing canonical scales $\mu_H = Q, \hspace{0.2cm} \mu_{J,B} = Q \ta^{1/(2-a)}, \hspace{0.2cm} \mu_S = Q \ta$, we can minimize the logarithms in each function at fixed-order in $\as$ then resum large logarithms by the RG evolution from the canonical scale to a common scale $\mu$ as in \eq{Gnu}. In small and perturbative region $\Lqcd/Q\ll \ta\ll 1$, the canonical choice works well. However, the scale profile as a function of $\ta$ should be properly modified from the canonical scales in the fixed-order region $\ta\sim \mathcal{O}(1)$ and the non-perturbative region $\ta\sim \Lqcd/Q$. For this purpose we divide the distribution into three regions 
\bea
&{\rm Peak~ region:}  & \mu_H \gg \mu_{J,B}  \gg  \mu_S \sim  \Lambda_{QCD}\nonumber \\
&{\rm Tail ~region:} & \mu_H \gg \mu_{J,B}  \gg  \mu_S \gg  \Lambda_{QCD}\\
&{\rm Far-tail~ region:} & \mu_H = \mu_{J,B}  =  \mu_S \gg  \Lambda_{QCD} \nonumber
\label{eq:regions}
\eea
In the tail region, the canonical scales work well. As we approach the peak region we needs to freeze the scales before soft scale reaches non-perturbative region, where our prediction is not valid and as we approach the far-tail region we make all scales merge the hard scale smoothly. 
The profile function satisfying these constraints has the following forms \cite{Bell:2018gce,Hoang:2014wka}
\bea 
\mu_H &=& e_H Q 
\nn \\
\mu_S(\ta) &=& \bigg[1+e_S \theta(t_3-\ta)\bigg(1- \frac{\ta}{t_3} \bigg)^2  \bigg] \mu_\text{run}(\ta) \label{eq:scales} \\
\mu_{J,B}(\ta) &=& \bigg[1+e_{J,B} \theta(t_3-\ta)\bigg(1- \frac{\ta}{t_3} \bigg)^2  \bigg] \mu_H^{\frac{1-a}{2-a}} \mu_\text{run}(\ta)^{\frac{1}{2-a}} 
\,, \nn
\eea
where the running scale $\mu_\text{run}$ is given by
\bea
\mu_\text{run}(\ta) = & \hspace{-4.5cm} \mu_0 & \ta \leq t_0 
\nonumber\\
=& \zeta(\ta; \{t_0,\mu_0,0\},\{t_1,0,\frac{r}{\ta^{sph}}\mu_H\}) & t_0 \leq \ta \leq t_1 \nonumber\\
=& \hspace{-3.5cm}  \frac{r}{\ta^{sph}}\mu_H \ta  & t_1 \leq \ta \leq t_2 \label{eq:prof} \\
=&  \zeta(\ta; \{t_2,0,\frac{r}{\ta^{sph}}\mu_H \},\{t_3,\mu_H,0 \}) & t_2 \leq \ta \leq t_3 \nonumber \\
=& \hspace{-4.5cm}  \mu_H & \ta \geq t_3. \nonumber
\eea
\begin{figure}[h]
\begin{center}
\includegraphics[scale=.3]{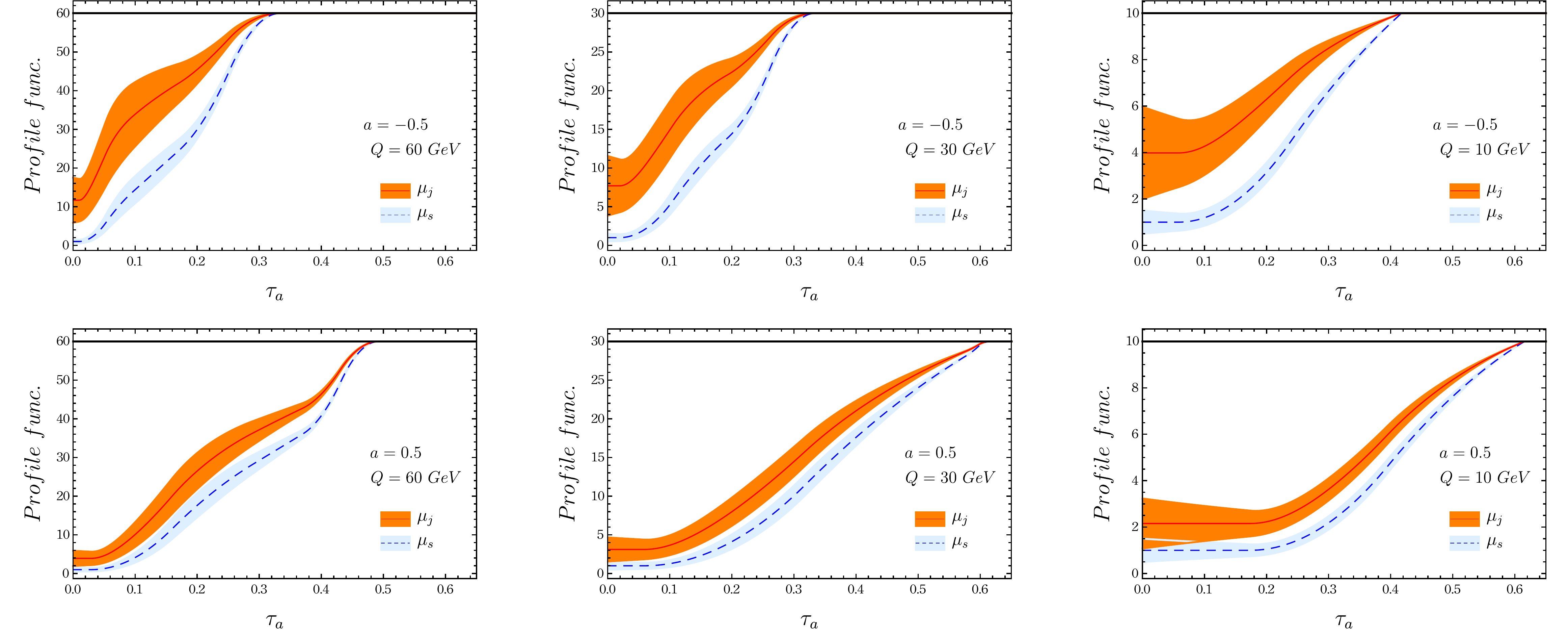} 
\end{center}
\caption{\label{fig:mprof_low} Modified profile functions for $Q=60, 30, 10\GeV$ at $a=-0.5, 0.5$. Red line with orange colored uncertainty represents $\mu_{J,B}$ and $\mu_S$ is shown in blue line with light-blue uncertainty region.}
\end{figure}
The $\zeta$ function is designed to ensure the continuity from initial region $\{t_i,y_i,r_1\}$ to final region $\{t_f,y_f,r_f\}$. The arguments $y_j$ and $r_j$ stands for the intercept and slope in that region.  The transition between the non-perturbative, resummation and fixed-order regions of the distributions are controlled by the parameters $t_j$ as
\bea 
t_0=& \frac{n_0}{Q}3^a, & \hspace{0.4cm} t_1= \frac{n_1}{Q}3^a ,\hspace{0.4cm} t_2= 0.85 \times 0.295^{1-0.637 a}, \hspace{0.4cm}t_3=0.8 \ta^{sph} , \label{eq:ti}
\eea
where $\ta^{sph}$ is the angularity of the spherically symmetric configuration at an arbitrary value of $a$ and at its maximum value $\ta^{max}=\tau^{sph}_{a^{max}}$.
For arbitrary $a$, $\tau_a^{sph}$ is given by:
\bea
\tau_{a}^{\mathrm{sph}}=\frac{1}{4 \pi} \int_{0}^{2 \pi} d \phi \int_{-1}^{1} d \cos \theta \sin ^{a} \theta(1-|\cos \theta|)^{1-a}=\frac{1}{2-\frac{a}{2}}{ }_{2} F_{1}\left(1,-\frac{a}{2} ; 3-\frac{a}{2} ;-1\right).
\eea
In this work, $\tau_{a}^{\mathrm{sph}}$ ranges from $\tau_{-1}^{\mathrm{sph}}\approx 0.356$ to $\tau_{0.5}^{\mathrm{sph}}\approx 0.616$.
The soft, jet and beam scales $\mu_S,\mu_{J,B}$ merge with the hard scale $\mu_H$ to have matching with the fixed-order perturbation theory in the far-tail region. 
The $a$-dependence in the $t_{0,1,2}$ are chosen based on the empirical observation, location of the peak scales as $3^a$ and $t_2$ is chosen as a point at which the singular and nonsingular contribution become equal in magnitude.  The values of $t_j$ would be different in DIS and needs to be determined by using an fixed-order result for DIS angularity. However it is absent in our paper and we use the same values used $e^+e^-$ angularity \cite{Bell:2018gce}. 

Theoretical uncertainty of our prediction is probed in band method \cite{Bell:2018gce}, where the parameters $n_{0,1}$ are taken as constant $n_0=1\GeV, n_1=10\GeV$.  We chose $n_0=1\GeV$ ensuring a comparable deviation from the canonical scale at point $t_0$.  In \eq{prof}, $r=1$ and $\mu_0=1\GeV$ and explicit form of the $\zeta$ function is given \cite{Hoang:2014wka,Bell:2018gce} as
\bea
\zeta(\ta; \{t_i,y_i,r_i \},\{t_f,y_f,r_f \}) 
=& 
\begin{cases}
a_i+r_i(\ta - t_i)+c_i(\ta - t_0)^2 
& \qquad \ta \leq \frac{t_i+t_f}{2} \,,\\
 a_f+r_f(\ta - t_f)+c_f(\ta - t_f)^2  
 &\qquad  \ta \geq \frac{t_i+t_f}{2}\,, 
\end{cases}
\eea
where the coefficients are
\bea 
a_i=& y_i+r_i t_i ,  \hspace{1cm}& c_i = 2\frac{a_f-a_i}{(t_i-t_f)^2}+ \frac{3r_i + r_f}{2(t_i-t_f)}\,,\\
a_f=& y_f+r_f t_f ,  \hspace{1cm} & c_f = 2\frac{a_i-a_f}{(t_i-t_f)^2}+ \frac{3r_f + r_i}{2(t_f-t_i)}
\eea

This profile function shows reasonable results at high $Q> 30\GeV$. According to the white paper,  EIC is going to cover a large range in $Q$ and we may need a access low $Q$ region experimentally. We noticed that this profile function shows discontinuity in $\mu_s, \mu_{J,B}$ for the low $Q$ and positive $a$ region. 
The reason for this discontinuity is as follows: from \eq{ti} one can easily see that at low energy  and positive $a$, $t_1$ becomes larger than $t_2,t_3$ and the $\zeta$-function in $\mu_\text{run}(\ta)$ failed to provide continuity at the transitions. Note that this discontinuity varies with $a$ as $t_1 \propto 3^a$ indicates, if $a$ increases the point of discontinuity moves towards large $\ta$ (one can easily see by plotting the profile functions for $Q\leq 30 \GeV$ and $a\sim  0.5$). 

To have access to the lower $Q$ region we modify the above profile function of \eq{prof} incorporating the following conditions: if $t_1 \geq t_2$, the $t_2$ is set to be equal to $t_1$ and so on for the $t_3$. The result for this modification is shown in \fig{mprof_low}, where first second and third columns are corresponding to $Q=60, 30, 10~ \GeV$ and two rows are for $a=-0.5, 0.5$. Each sub-plots represents the modified jet scale $\mu_J$ in orange continuous line and soft scale $\mu_S$ in blue dashed line. The colored (orange and blue) bands represent uncertainty in the corresponding scale variation. These scale variation is operated by varying  $e_H=2^{\pm 1}, e_{S}= \pm 1/2$ and $e_{J,B}= \pm 1/2$ in \eq{scales}. To have a better visibility, we do not show uncertainty in the hard scale as it simply varies with single power in $Q$. 
\fig{mprof_low} shows that our modification in profile function provides access to minimum scale $Q_\text{min} \sim 10\GeV $ and $a=0.5$. 

\section{Anomalous dimensions and related integral} \label{app:anom}

Here, we summarize our convention for logarithmic accuracy and give the expressions of anomalous dimensions and beta function coefficient used in the resummation.

Solution of RGE in \eq{dGdmu} contains the exponent of integrated anomalous dimensions as shown in \eq{intgam}, which essentially resums logarithms. 
Here, our power counting for a large log $L$ is $\as L\sim \cO(1)$ and leading log is a term like $\as^n L^{n+1}$ which is of order $ L\sim 1/\as$.
Then, next-to-leading log (NLL) is like $\as^n L^{n}$ and
N$^k$LL is like $\as^{n+k}L^{n}$.
With this log counting, if we 
insert 1-loop cusp result $\propto\as \Gamma_ 0$ into \eq{intgam}, we can find that
the leading log term $\as \ln^2(\mu/\mu_G)$ at order $\as$ is obtained without $\as$ evolution and 
all leading log terms are captured correctly
with $\as$ evolution with 1-loop beta function coefficients $\propto \as \beta_0$. In RGE, the non-cusp term is suppressed by one power of log and begin to contribute from NLL accuracy. The fixed-order functions with no large log have ordinary $\as$ counting then, $\as^n$ term is comparable with N$^{n+1}$LL accuracy. \tab{logaccuracy} summarizes ingredients needed at each log accuracy.

\begin{table}[t]
 $$
\begin{array}{|c|c|c|c|c|}
\hline
 & \Gamma(\as) & \gamma(\as) & \beta(\as) & \{H,J,B,S\}[\as] \\ \hline
 \text{LL} & \as & 1 & \as & 1 \\ \hline
  \text{NLL} & \as^2 & \as & \as^2 & 1 \\ \hline
    \text{NNLL} & \as^3 & \as^2 & \as^3 & \as \\ \hline
\end{array}
 $$
\caption{
Resummation accuracy and corresponding order of individual ingredients: cusp and non-cusp anomalous dimensions, beta function, and fixed-order hard, jet, beam, soft functions
\label{tab:logaccuracy}}
\end{table}

To the NNLL order we need~\cite{Korchemsky:1987wg, Moch:2004pa},
\begin{subequations}
\begin{align}
\Gamma_0 &= 4 C_F \nn\\
\Gamma_1 &= \Gamma_0 \Bigl[ \Bigl( \frac{67}{9} -\frac{\pi^2}{3}\Bigr) C_A - \frac{20}{9} T_F n_f\Bigr] \nn\\
\Gamma_2 &= \Gamma_0 \Bigl[ \Bigl( \frac{245}{6} - \frac{134\pi^2}{27} + \frac{11\pi^4}{45} + \frac{22\zeta_3}{3} \Bigr) C_A^2 + \Bigl( - \frac{418}{27} + \frac{40\pi^2}{27} - \frac{56\zeta_3}{3}\Bigr) C_A T_F n_f \nn\\
&\qquad + \Bigl(-\frac{55}{3} + 16\zeta_3\Bigr) C_F T_F n_f - \frac{16}{27} T_F^2 n_f^2\Bigr]
\end{align}
\end{subequations}

The beta function expanded in powers of $\alpha_s$ is given by
\be
\beta(\alpha_s) =\mu \frac{d \as(\mu)}{d\mu}
=- 2 \alpha_s \sum_{n=0}^\infty \beta_n\Bigl(\frac{\alpha_s}{4\pi}\Bigr)^{n+1}
\ee
The coefficients~\cite{Tarasov:1980au, Larin:1993tp} are 
\begin{align} \label{eq:betai}
\beta_0 &= \frac{11}{3}\,C_A -\frac{4}{3}\,T_F\,n_f
\,,\nn\\
\beta_1 &= \frac{34}{3}\,C_A^2  - \Bigl(\frac{20}{3}\,C_A\, + 4 C_F\Bigr)\, T_F\,n_f
\,, \nn\\
\beta_2 &=
\frac{2857}{54}\,C_A^3 + \Bigl(C_F^2 - \frac{205}{18}\,C_F C_A
 - \frac{1415}{54}\,C_A^2 \Bigr)\, 2T_F\,n_f
 + \Bigl(\frac{11}{9}\, C_F + \frac{79}{54}\, C_A \Bigr)\, 4T_F^2\,n_f^2
\,.\end{align}

Integrated anomalous dimensions are defined in \eq{Keta-def} and their explicit expressions in powers of $\as$ are
\begin{align} 
K_\Gamma(\mu_0, \mu) &= -\frac{\Gamma_0}{4\beta_0^2}\,
\biggl\{ \frac{4\pi}{\alpha_s(\mu_0)}\, \Bigl(1 - \frac{1}{r} - \ln r\Bigr)
   + \biggl(\frac{\Gamma_1 }{\Gamma_0 } - \frac{\beta_1}{\beta_0}\biggr) (1-r+\ln r)
   + \frac{\beta_1}{2\beta_0} \ln^2 r
\nn\\ & \quad
+ \frac{\alpha_s(\mu_0)}{4\pi}\, \biggl[
  \biggl(\frac{\beta_1^2}{\beta_0^2} - \frac{\beta_2}{\beta_0} \biggr) \Bigl(\frac{1 - r^2}{2} + \ln r\Bigr)
  + \biggl(\frac{\beta_1\Gamma_1 }{\beta_0 \Gamma_0 } - \frac{\beta_1^2}{\beta_0^2} \biggr) (1- r+ r\ln r)
 \nn\\ & \hspace{3cm}
  - \biggl(\frac{\Gamma_2 }{\Gamma_0} - \frac{\beta_1\Gamma_1}{\beta_0\Gamma_0} \biggr) \frac{(1- r)^2}{2}
     \biggr] \biggr\}
\,, \nn\\
\eta_\Gamma(\mu_0, \mu) &=
 - \frac{\Gamma_0}{2\beta_0}\, \biggl[ \ln r
 + \frac{\alpha_s(\mu_0)}{4\pi}\, \biggl(\frac{\Gamma_1 }{\Gamma_0 }
 - \frac{\beta_1}{\beta_0}\biggr)(r-1)
 + \frac{\alpha_s^2(\mu_0)}{16\pi^2} \biggl(
    \frac{\Gamma_2 }{\Gamma_0 } - \frac{\beta_1\Gamma_1 }{\beta_0 \Gamma_0 }
      + \frac{\beta_1^2}{\beta_0^2} -\frac{\beta_2}{\beta_0} \biggr) \frac{r^2-1}{2}
    \biggr]
\,, \nn\\
K_\gamma(\mu_0, \mu) &=
 - \frac{\gamma_0}{2\beta_0}\, \biggl[ \ln r
 + \frac{\alpha_s(\mu_0)}{4\pi}\, \biggl(\frac{\gamma_1 }{\gamma_0 }
 - \frac{\beta_1}{\beta_0}\biggr)(r-1) \biggr]
\label{eq:Keta}
\,.\end{align}
Here, $r = \alpha_s(\mu)/\alpha_s(\mu_0)$. Solving the beta function to three-loop order gives the running coupling expressed by
\begin{align} \label{eq:alphas}
\frac{1}{\alpha_s(\mu)} &= \frac{X}{\alpha_s(\mu_0)}
  +\frac{\beta_1}{4\pi\beta_0}  \ln X
  + \frac{\alpha_s(\mu_0)}{16\pi^2} \biggr[
  \frac{\beta_2}{\beta_0} \Bigl(1-\frac{1}{X}\Bigr)
  + \frac{\beta_1^2}{\beta_0^2} \Bigl( \frac{\ln X}{X} +\frac{1}{X} -1\Bigr) \biggl]
\,,\end{align}
where $X\equiv 1+\alpha_s(\mu_0)\beta_0 \ln(\mu/\mu_0)/(2\pi)$.
In our numerical calculations we take the full NNLL results in \eq{Keta} for $K_{\Gamma,\gamma},\eta_\Gamma$ and in \eq{alphas}. To be consistent with the value of $\as(\mu)$ we take the NNLO PDFs in our numerical results.

\bibliography{angDIS}
\end{document}